# Continuously Guided Atomic Interferometry Using a Single-Zone Optical Excitation: Theoretical Analysis

M.S. Shahriar [1,2], M. Jheeta [2], Y. Tan [2], P. Pradhan[1,2] and A. Gangat[1]

[1]*Deptartment of Electrical and Computer Engineering, Northwestern University Evanston, IL 60208*

[2]*Research Laboratory of Electronics, Massachusetts Institute of Technology Cambridge, MA 02139*

**Abstract**

In an atomic interferometer, the phase shift due to rotation is proportional to the area enclosed by the split components of the atom. However, this model is unclear for an atomic interferometer demonstrated recently by Shahriar *et al.*, for which the atom simply passes through a *single-zone* optical beam, consisting of a pair of bichromatic counter-propagating beams. During the passage, the atomic wave packets in two distinct internal states couple to each other *continuously*. The two internal states trace out a complicated trajectory, guided by the optical beams, with the amplitude and spread of each wavepacket varying continuously. Yet, at the end of the single-zone excitation, there is an interference with fringe amplitudes that can reach a visibility close to unity. For such a situation, it is not clear how one would define the area of the interferometer, and therefore, what the rotation sensitivity of such an interferometer would be. In this paper we analyze this interferometer in order to determine its rotation sensitivity, and thereby determine its effective area. In many ways, the continuous interferometer (CI) can be thought of as a limiting version of the Borde-Chu Interferometer (BCI). We identify a quality factor that can be used to compare the performance of these interferometers. Under conditions of practical interest, we show that the rotation sensitivity of the CI can be comparable to that of the BCI. The relative simplicity of the CI (e.g., elimination of the task of precise angular alignment of the three zones) then makes it a potentially better candidate for practical atom interferometry for rotation sensing.

## 1. Introduction

In an atomic interferometer[1-8], the phase shift due to rotation is proportional to the area enclosed by the split components of the atom. In most situations, the atomic wavepacket is split first by what can be considered effectively as an atomic beam-splitter [9-13]. The split components are then redirected towards each other by atomic mirrors. Finally, the converging components are recombined by another atomic beam splitter. Under these conditions, it is simple to define the area of the interferometer by considering the center of mass motion of the split components. However, this model is invalid for an atomic interferometer demonstrated recently by Shahriar *et al.* [14]. Briefly, in this interferometer, the atom simply passes through a *single-zone* optical beam, consisting of a pair of bichromatic counter-propagating beams. During the passage, the atomic wave packets in two distinct internal states couple to each other *continuously*. The two internal states trace out complicated trajectories, guided by the optical beams, with the amplitude and spread of each wavepacket varying continuously. Yet, at the end of the single-zone excitation, there is an interference with fringe amplitudes that can reach a visibility close to unity. For such a situation, it is not clear how one would define the area of the interferometer, and therefore, what the rotation sensitivity of such an interferometer would be.

In this paper we analyze this interferometer in order to determine its rotation sensitivity, and thereby determine its effective area. In many ways, the continuous interferometer (CI) can be thought of as a limiting version of the three-zone interferometer proposed originally by Borde [1], and demonstrated by Chu *et al.* [2]. In our analysis, we compare the behavior of the CI with the Borde-Chu Interferometer (BCI). We also identify a quality factor that can be used to compare the performance of these interferometers. Under conditions of practical interest, we show that the rotation sensitivity of the CI can be comparable to that of the BCI. The relative simplicity of the CI (e.g., the task of precise angular alignment of the three zones is eliminated for the CI) then makes it a potentially better candidate for practical atom interferometry for rotation sensing.

In our comparative analysis, we find it more convenient to generalize the BCI by making the position and duration of the phase-scanner a variable. As such, we end up comparing two types of atomic interferometers to the BCI. The first, which is the generalized version of the BCI, is where instead of a phase scan being applied in only the final $\pi/2$ pulse, the phase scan is applied from some point onwards in the middle $\pi$ pulse. We find that the magnitude of the rotational phase shift varies according to where the phase is applied from. This phase shift is calculated analytically and compared to the phase shift obtained in the original BCI. The second type of interferometer that we compare to the original BCI is the CI, where the atom propagates through only one laser beam that has a Gaussian field profile. The atom is modeled as a wavepacket with a Gaussian distribution in the momentum representation, and it's evolution in the laser field is calculated numerically. From this, the rotational phase shift is obtained and compared once again to the phase shift in the BCI.

## 2. Formulation of The Problem

We model the system as a three level atom in the lambda configuration, as shown in Fig. 1, with levels |a>, |b>, and |e>, which moves in the **x** direction through two counter propagating laser beams. The laser beams travel in the **z** direction and have Gaussian electric field profiles varying in the **x** direction. In the electric dipole approximation, which is valid for our system since the wavelength of the light is much greater than the separation between the electron and the nucleus, we can write the interaction Hamiltonian as **r**.**E**, where **r** is the position of the electron and **E** is the electric field of the laser. The states of the three level atom are driven by the laser fields. The fields cause transitions between the states |a> and |e> and the states |e> and |b>. In our analysis, we quantize the center-of-mass (COM) position of the atom in the z direction. The Hamiltonian for the system can be written in the following way:

$$H = \frac{\mathbf{P}_{\mathbf{z}}^{\mathbf{2}}}{2m} + H_0 + \mathbf{r} \cdot \mathbf{E}_{\mathbf{1}} + \mathbf{r} \cdot \mathbf{E}_{\mathbf{2}} \tag{1}$$

where $\mathbf{E}_1$ and $\mathbf{E}_2$ are the electric field vectors of the two counter propagating lasers, $\mathbf{P_z}$ is the COM momentum in the **z** direction, and $H_o$ is the internal energy. The lasers are taken to be classical electromagnetic fields. We can expand the Hamiltonian as well as the wavefunction of the COM in the basis of the eigenstates of the non-interacting Hamiltonian, which is simply: $|P_Z\rangle \otimes |i\rangle = |P_Z, i\rangle$. This is a complete set of basis states for our system. Since it is understood that all momenta and positions refer to the **z** direction, we will drop the z subscript on all momenta from hereon.

The position operator of the electron in the atom can be expanded in terms of this basis by inserting the identity operator twice in the form

$$\hat{I} = \int dp \sum_i |p,i\rangle\langle p,i| \quad . \tag{2}$$

We also make the assumption that matrix elements of the form $\langle i| \mathbf{r} |i\rangle = 0$. Thus, in terms of the dipole matrix elements $\mathbf{d}_{ij} = \langle i| \mathbf{r} |j\rangle$, we can write the position operator as:

$$\begin{aligned} \mathbf{r} &= \int dp \int dp' \sum_{i,j} |p,i\rangle\langle p,i| \mathbf{r} |p',j\rangle\langle p',j| \\ &= \int dp \left( \mathbf{d}_{ae} |p,a\rangle\langle p,e| + \mathbf{d}_{ea} |p,e\rangle\langle p,a| + \mathbf{d}_{be} |p,b\rangle\langle p,e| + \mathbf{d}_{eb} |p,e\rangle\langle p,b| \right) . \end{aligned} \tag{3}$$

Define the atomic raising and lowering operators as:

$$\sigma_{ij} = |p,i\rangle\langle p,j| \quad . \tag{4}$$

In terms of these operators, the position operator is

$$\mathbf{r} = \int dp \left( \mathbf{d}_{ae}\sigma_{ae} + \mathbf{d}_{ea}\sigma_{ea} + \mathbf{d}_{be}\sigma_{be} + \mathbf{d}_{eb}\sigma_{eb} \right). \tag{5}$$

Since the electric field is being modeled classically, we can express each laser field as

$$\mathbf{E}(z,t) = \mathbf{E}_0 \cos(\omega t - k\hat{z} + \phi) = \frac{\mathbf{E}_0}{\mathbf{2}}\left(\exp(i(\omega t - k\hat{z} + \phi)) + \exp(-i(\omega t - k\hat{z} + \phi))\right), \tag{6}$$

where $\hat{z}$ is the operator associated with the COM position of the atom. The first laser interacts only with that part of the electron position operator which causes transitions |a> ↔ |e>, and the second laser interacts with the part which causes transitions |b> ↔ |e>. We also assume that the dipole matrix elements are real: $\mathbf{d}_{ae} = \mathbf{d}_{ea} = \mathbf{d}_{ae}^*$. Therefore,

$$\begin{aligned}\mathbf{r}\cdot\mathbf{E}_1 = \int dp\, \frac{\mathbf{d}_{ae}\cdot\mathbf{E}_{01}}{2}\big[&\sigma_{ae}\exp\left(i\left(\omega_1 t - k_1\hat{z} + \phi_1\right)\right) + \sigma_{ae}\exp\left(-i\left(\omega_1 t - k_1\hat{z} + \phi_1\right)\right) \\ &+ \sigma_{ea}\exp\left(i\left(\omega_1 t - k_1\hat{z} + \phi_1\right)\right) + \sigma_{ea}\exp\left(-i\left(\omega_1 t - k_1\hat{z} + \phi_1\right)\right)\big]\end{aligned} \tag{7}$$

Now we make the standard rotating wave approximation which neglects the terms in this expression which do not conserve energy. Also, let $\Omega_1 = \frac{\mathbf{d}_{ae}\cdot\mathbf{E}_{01}}{\hbar}$, so that

$$\mathbf{r}\cdot\mathbf{E}_1 = \int dp\, \frac{\hbar\Omega_1}{2}\left[\sigma_{ae}\exp\left(i\left(\omega_1 t - k_1\hat{z} + \phi_1\right)\right) + \sigma_{ea}\exp\left(-i\left(\omega_1 t - k_1\hat{z} + \phi_1\right)\right)\right], \tag{8}$$

with a corresponding expression for the **r·E₂** part. Now, we get:

$$\begin{aligned}\exp(ik\hat{z}) &= \sum_{i,j}\int dp \int dp' \left|p,i\right\rangle\left\langle p,i\right|e^{ikz}\left|p',j\right\rangle\left\langle p',j\right| \\ &= \sum_{i,j}\int dp\int dp'\int dz\int dz' \left|p,i\right\rangle\left\langle p,i\middle|z,i\right\rangle\left\langle z,i\right|e^{ikz}\left|z',j\right\rangle\left\langle z',j\middle|p',j\right\rangle\left\langle p',j\right| \\ &= \sum_{i,j}\int dp\int dp'\int dz\int dz' \left|p,i\right\rangle e^{\frac{-ipz}{\hbar}} e^{ikz'} e^{\frac{ip'z'}{\hbar}} \delta_{ij}\delta(z-z')\left\langle p',j\right| \\ &= \sum_{i}\int dp\int dp' \left|p,i\right\rangle\left\langle p',j\right|\delta\!\left(k - \frac{p}{\hbar} + \frac{p'}{\hbar}\right) \\ &= \sum_{i}\int dp \left|p,i\right\rangle\left\langle p - \hbar k, i\right|\end{aligned} \tag{9}$$

and

$$\exp(-ik\hat{z}) = \sum_{i}\int dp \left|p,i\right\rangle\left\langle p + \hbar k, i\right|. \tag{10}$$

The expansion of the non-interacting part of the Hamiltonian in Eq. (1) in terms of this basis is

$$H_0 = \int dp \sum_i \left( \frac{p^2}{2m} + \hbar\omega_i \right) \sigma_{ii} \ , \tag{11}$$

where $\hbar\omega_i$ is the energy of the i-th level.

Combining expressions of Eqs. (7)-(11) in Eq. (1), we finally get the full Hamiltonian in the |p,i> basis:

$$\begin{aligned} H = \int dp \Bigg[ \sum_i \left( \frac{p^2}{2m} + \hbar\omega_i \right) \sigma_{ii} &+ \frac{\hbar\Omega_1}{2} \left( \left| p,a \right\rangle \left\langle p + \hbar k_1, e \right| e^{i(\omega_1 t + \phi_1)} + \left| p + \hbar k_1, e \right\rangle \left\langle p,a \right| e^{-i(\omega_1 t + \phi_1)} \right) \\ &+ \frac{\hbar\Omega_2}{2} \left( \left| p,b \right\rangle \left\langle p + \hbar k_2, e \right| e^{i(\omega_2 t + \phi_2)} + \left| p + \hbar k_2, e \right\rangle \left\langle p,b \right| e^{-i(\omega_2 t + \phi_2)} \right) \Bigg]. \end{aligned} \tag{12}$$

For a given value of the momentum p, it is clear that this Hamiltonian creates transitions only between the following manifold of states, $\left| p,a \right\rangle \leftrightarrow \left| p + \hbar k_1, e \right\rangle \leftrightarrow \left| p + \hbar k_1 - \hbar k_2, b \right\rangle$. That is, the only way to transition between states $\left| a \right\rangle$ and $\left| b \right\rangle$ is to pass through $\left| e \right\rangle$ and make the accompanying momentum transitions as indicated. Therefore it is convenient to make the following substitutions of the momentum variables:

$$\int dp \left( \frac{p^2}{2m} + \hbar\omega_e \right) \left| p,e \right\rangle \left\langle p,e \right| = \int dq_1 \left( \frac{(q_1 + \hbar k_1)^2}{2m} + \hbar\omega_e \right) \left| q_1 + \hbar k_1, e \right\rangle \left\langle q_1 + \hbar k_1, e \right|, \tag{13a}$$

$$\begin{aligned} &\int dp \left( \frac{p^2}{2m} + \hbar\omega_b \right) \left| p,b \right\rangle \left\langle p,b \right| = \\ &\qquad \int dq_2 \left( \frac{(q_2 + \hbar k_1 - \hbar k_2)^2}{2m} + \hbar\omega_b \right) \left| q_2 + \hbar k_1 - \hbar k_2, b \right\rangle \left\langle q_2 + \hbar k_1 - \hbar k_2, b \right|, \end{aligned} \tag{13b}$$

$$\int dp \left| p,b \right\rangle \left\langle p + \hbar k_2, e \right| = \int dq_2 \left| q_2 + \hbar k_1 - \hbar k_2, b \right\rangle \left\langle q_2 + \hbar k_1, e \right|. \tag{13c}$$

Thus if we define the states

$$\left| 1 \right\rangle = \left| p,a \right\rangle, \tag{14a}$$

$$\left| 2 \right\rangle = \left| p + \hbar k_1, e \right\rangle, \tag{14b}$$

$$\left| 3 \right\rangle = \left| p + \hbar k_1 - \hbar k_2, b \right\rangle, \tag{14c}$$

we can rewrite the Hamiltonian (Eq. (12) ) as

$$H = \int dp \left[ \sum_i \varepsilon_i \sigma_{ii} + \frac{\hbar\Omega_1}{2}\left( |1\rangle\langle 2| e^{i(\omega_1 t + \phi_1)} + |2\rangle\langle 1| e^{-i(\omega_1 t + \phi_1)} \right) + \right.$$
$$\left. + \frac{\hbar\Omega_2}{2}\left( |3\rangle\langle 2| e^{i(\omega_2 t + \phi_2)} + |2\rangle\langle 3| e^{-i(\omega_2 t + \phi_2)} \right) \right], \tag{15}$$

where the $\varepsilon_i$ are the energies of the newly defined states. This form of the Hamiltonian makes it clear that once the atom has some momentum p, the interaction cannot move it to a manifold of states with some other momentum. The only transitions that can occur are between the states |1>, |2> and |3>, for the given momentum. Thus, to study the dynamics of the atom, it is sufficient to consider only one manifold with some momentum p. Once solved, we can integrate over all momenta to get the motion of the full wavepacket.

Since the laser beams are counter-propagating at the same frequency, we have $k_1 = - k_2 = k$, and the states become:

$$|1\rangle = |p, a\rangle, \tag{16a}$$
$$|2\rangle = |p + \hbar k, e\rangle, \tag{16b}$$
$$|3\rangle = |p + 2\hbar k, b\rangle. \tag{16c}$$

The state of the atom is expanded in the |p,i> basis as

$$|\Psi(t)\rangle = \int dp \sum_i \psi_i(p,t) |p,i\rangle$$
$$= \int dp \left( \alpha(p,t) |p,a\rangle + \beta(p + 2\hbar k, t) |p + 2\hbar k, b\rangle + \xi(p + \hbar k, t) |p + \hbar k, e\rangle \right), \tag{17}$$

and evolves according to the Schreodinger equation:

$$i\hbar \frac{\partial |\Psi\rangle}{\partial t} = H\Psi \,. \tag{18}$$

If we make a unitary transformation U on the state $|\Psi\rangle$ to some interaction-picture state vector $|\tilde{\Psi}\rangle = U|\Psi\rangle$, then the Hamiltonian in this interaction picture is

$$\tilde{H} = UHU^{-1} + i\hbar \frac{\partial U}{\partial t} U^{-1} \,. \tag{19}$$

Let $U = \int dp \sum_j e^{i(\theta_j t + \varsigma_j)} |j\rangle$, where the |j> are the redefined states of Eq. (16), and the $\theta_j$ and $\varsigma_j$ are parameters we will choose to simplify the interaction picture Hamiltonian. Written in matrix form, the Hamiltonian for some momentum p is

$$H(p)=\begin{bmatrix} \varepsilon_1 & 0 & \frac{\hbar\Omega_1}{2}e^{i(\omega_1 t+\phi_1)} \\ 0 & \varepsilon_3 & \frac{\hbar\Omega_2}{2}e^{i(\omega_2 t+\phi_2)} \\ \frac{\hbar\Omega_1}{2}e^{-i(\omega_1 t+\phi_1)} & \frac{\hbar\Omega_2}{2}e^{-i(\omega_2 t+\phi_2)} & \varepsilon_2 \end{bmatrix}, \tag{20}$$

where the rows and columns are arranged with the states in the {|1>, |3>, |2>} order. In the interaction picture with the parameters $\theta_j$ and $\varsigma_j$, the Hamiltonian is

$$\widetilde{H}(p)=\begin{bmatrix} \varepsilon_1-\hbar\theta_1 & 0 & \frac{\hbar\Omega_1}{2}e^{i(\omega_1+\theta_1-\theta_2)t+i(\phi_1+\varsigma_1-\varsigma_2)} \\ 0 & \varepsilon_3-\hbar\theta_3 & \frac{\hbar\Omega_2}{2}e^{i(\omega_2+\theta_3-\theta_2)t+i(\phi_2+\varsigma_3-\varsigma_2)} \\ \frac{\hbar\Omega_1}{2}e^{-i(\omega_1+\theta_1-\theta_2)t-i(\phi_1+\varsigma_1-\varsigma_2)} & \frac{\hbar\Omega_2}{2}e^{-i(\omega_2+\theta_3-\theta_2)t-i(\phi_2+\varsigma_3-\varsigma_2)} & \varepsilon_2-\hbar\theta_2 \end{bmatrix}. \tag{21}$$

First, to get rid of the time dependence, set $\omega_1+\theta_1-\theta_2=0$, and $\omega_2+\theta_3-\theta_2=0$. Also, set $\varsigma_1=-\phi_1$, $\varsigma_2=0$, and $\varsigma_3=-\phi_2$. Define the detunings:

$$\hbar\delta_1=\varepsilon_1+\hbar\omega_1-\varepsilon_2, \tag{22a}$$
$$\hbar\delta_2=\varepsilon_3+\hbar\omega_2-\varepsilon_2, \tag{22b}$$
$$\hbar\Delta=\hbar(\delta_1-\delta_2), \tag{22c}$$
$$\hbar\delta=\frac{\hbar(\delta_1+\delta_2)}{2}. \tag{22d}$$

A consistent choice of the θ parameters that also simplifies the form of the Hamiltonian considerably is:

$$\hbar\theta_1=\frac{\varepsilon_1+\varepsilon_3-\hbar\omega_1+\hbar\omega_2}{2}, \tag{23a}$$
$$\hbar\theta_2=\frac{\varepsilon_1+\varepsilon_3+\hbar\omega_1+\hbar\omega_2}{2}, \tag{23b}$$
$$\hbar\theta_3=\frac{\varepsilon_1+\varepsilon_3+\hbar\omega_1-\hbar\omega_2}{2}. \tag{23c}$$

With this choice, the Hamiltonian becomes

$$\tilde{H}(p) = \hbar \begin{bmatrix} \frac{\Delta}{2} & 0 & \frac{\Omega_1}{2} \\ 0 & -\frac{\Delta}{2} & \frac{\Omega_2}{2} \\ \frac{\Omega_1}{2} & \frac{\Omega_2}{2} & -\delta \end{bmatrix}. \tag{24}$$

which is familiar from semi-classical (i.e., without quantization of the COM motion) descriptions of the three-level interaction [15-19], keeping in mind that here it represents the Hamiltonian only within a given manifold. The equations of motion for these three states within the manifold of a given momentum are:

$$i\hbar\dot{\tilde{\alpha}}(p,t) = \frac{\hbar\Delta}{2}\tilde{\alpha}(p,t) + \frac{\hbar\Omega_1}{2}\tilde{\xi}(p+\hbar k,t), \tag{25a}$$

$$i\hbar\dot{\tilde{\beta}}(p+2\hbar k,t) = -\frac{\hbar\Delta}{2}\tilde{\beta}(p+2\hbar k,t) + \frac{\hbar\Omega_2}{2}\tilde{\xi}(p+\hbar k,t), \tag{25b}$$

$$i\hbar\dot{\tilde{\xi}}(p+\hbar k,t) = -\hbar\delta\,\tilde{\xi}(p+\hbar k,t) + \frac{\hbar\Omega_1}{2}\tilde{\alpha}(p,t) + \frac{\hbar\Omega_2}{2}\tilde{\beta}(p+2\hbar k,t). \tag{25c}$$

Since the laser beams are far detuned from resonances, we make the adiabatic approximation, which can be verified afterwards for consistency. This approximation assumes that the intermediate |2> state occupation is negligible and that we can set $\dot{\tilde{\xi}} \approx 0$. This allows us to reduce this three level system to a two level system by solving for $\tilde{\xi}$ in Eq. (25c), and then substituting it into Eqs. (25a) and (25b). We get

$$\tilde{\xi} = \frac{\Omega_1}{2\delta}\tilde{\alpha} + \frac{\Omega_2}{2\delta}\tilde{\beta}, \tag{26}$$

and the effective two level Hamiltonian

$$\tilde{H}_{eff}(p) = \hbar \begin{bmatrix} \frac{\Delta}{2} + \frac{\Omega_1^2}{4\delta} & \frac{\Omega_1\Omega_2}{4\delta} \\ \frac{\Omega_1\Omega_2}{4\delta} & -\frac{\Delta}{2} + \frac{\Omega_2^2}{4\delta} \end{bmatrix}. \tag{27}$$

In our system, we assume that the counter-propagating laser beams have the same strength, $\Omega_1 = \Omega_2$, and define the effective Raman Rabi frequency $\Omega_o = \frac{\Omega_1\Omega_2}{2\delta}$. Thus the effective Hamiltonian becomes

$$\widetilde{H}_{eff}(p) = \hbar \begin{bmatrix} \frac{\Delta}{2} + \frac{\Omega_o}{2} & \frac{\Omega_o}{2} \\ \frac{\Omega_o}{2} & -\frac{\Delta}{2} + \frac{\Omega_o}{2} \end{bmatrix}. \tag{28}$$

The expressions for the detunings are:

$$\Delta = \Delta_0 - \frac{2kp}{m} - \frac{2\hbar k^2}{m}, \tag{29a}$$

$$\delta = \delta_0 + \frac{\hbar k^2}{2m}, \tag{29b}$$

where $\Delta_0 = \omega_1 - \omega_2 + \omega_a - \omega_b$ and $\delta_0 = (\omega_1 + \omega_2 + \omega_a + \omega_b - 2\omega_e)/2$. This effective Hamiltonian can be solved by standard methods for $\widetilde{\alpha}(p,t)$ and $\widetilde{\beta}(p + 2\hbar k, t)$. Once we have the solutions for some $\Omega_o$ and at a given value of p, then we can write down the full expression for the state vector integrated over all p. Ignoring any global phase factors which do not depend on p, we get

$$\begin{aligned} |\Psi(t)\rangle &= \int dp\, \left(\alpha(p,t)|p,a\rangle + \beta(p+2\hbar k,t)|p+2\hbar k,b\rangle\right) \\ &= \int dp\, \left(e^{-i\theta_1 t}\widetilde{\alpha}(p,t)|p,a\rangle + e^{-i\theta_3 t}\widetilde{\beta}(p+2\hbar k,t)|p+2\hbar k,b\rangle\right) \\ &= \int dp\, \exp\left(\frac{-i\left(p^2 + (p+2\hbar k)^2\right)}{4m\hbar}t\right)\left(\widetilde{\alpha}(p,t)|p,a\rangle + \widetilde{\beta}(p+2\hbar k,t)|p+2\hbar k,b\rangle\right). \end{aligned} \tag{30}$$

In our analysis of the rotational sensitivity, we must apply this solution for the state vector for the case of a Gaussian profile in the x direction. We simply discretize the Gaussian profile and propagate stepwise along the discrete profile until we reach the time desired. The position representation of the wavefunctions for the |a> and |b> states are then:

$$\psi_a(x,t) = \int dp\, \alpha(p,t)\exp(\frac{-ipx}{\hbar}), \tag{31a}$$

$$\psi_b(x,t) = \int dp\, \beta(p+2\hbar k,t)\exp(\frac{-ipx}{\hbar}), \tag{31b}$$

and the probabilities for the atom to be in either state are:

$$P(a) = \int dp\, |\alpha(p,t)|^2, \tag{32a}$$

$$P(b) = \int dp\, |\beta(p,t)|^2. \tag{32b}$$

## 3. Rotational Sensitivity

In the setup for the Borde-Chu Interferometer, shown in Fig. 2, where we assume that the transverse displacement is negligible as compared to the longitudinal travel (L>>d), the phase shift due to rotation of the interferometer may be interpreted as resulting from the deviation in position of the laser beams with respect to the atomic trajectories. This can be seen as follows. When the BCI is stationary, the laser fields may be assumed, without loss of generality, not to have any phase difference relative to one another. Once the BCI begins rotating with some angular velocity $\Omega$ around an arbitrary axis, each of the laser beams will move a distance relative to the axis of rotation in proportion to $\Omega$. This deviation in position results in a phase shift in each of the laser fields, given by $2\,k\,\Delta y$, where k is the wave number of the lasers and $\Delta y$ is the change in position. The factor of 2 results from the fact that two counter-propagating beams are used in each zone. The total phase shift due to the lasers in the BCI is $\delta\phi = \phi_1 - 2\phi_2 + \phi_3$, where $\phi_i\,(i = 1,2,3)$ corresponds to the phase deviation of the i$^{th}$ laser field. We assume, without loss of generality, that $\phi_i = 0 (i = 1,2,3)$ in the absence of any rotation, or any external phase shift applied to the beams. The rotational phase $\delta\phi_0$ is calculated by taking into account the phase shift of each laser field resulting from its respective change in position, by the time the atom reaches that field. If we choose the interferometer to rotate around point A, for example, then $\Delta y_1 = 0$, $\Delta y_2 = L\Omega T$, $\Delta y_3 = 4L\Omega T$, where T is the time the atom takes to go between lasers. Thus the phases associate with the rotation of the interferometer: $\phi_{10} = 0$, $\phi_{20} = 2kL\Omega T$, $\phi_{30} = 8kL\Omega T$, and $\delta\phi_0 = 4kL\Omega T$. Since $v_y = 2\hbar k / m$ (from photon recoil), $d = v_y T$, and the area of the interferometer is $A_o = Ld = 2LT\hbar k/m$, we get

$$\delta\phi_0 = 4kL\Omega T = 2\Omega A_o m / \hbar \ . \tag{33}$$

This expression for the rotational phase remains the same regardless of the position of the axis of rotation, as can be shown explicitly. We are neglecting second order contributions to the rotational phase, which come from the difference in path lengths between the upper and lower arms while rotating. Of course, even though this expression is derived here explicitly for the BCI, it is in fact applicable to any atomic interferometer as long as the area enclosed (as defined by the semi-classical trajectories) is given by $A_o$ [20, 21]. The expression can also be derived from the corresponding expression for the rotation sensitivity of an optical gyroscope [$4\pi\Omega A_o/\lambda c$] based on the Sagnac effect [22], by substituting $mc^2$ (the rest energy of the atom) for $h\nu$, the photon energy.

During a typical operation of the BCI [2-4], an additional phase shift $\phi$ is applied on the third zone of laser beams so that $\delta\phi = \phi + \delta\phi_0$. As the value of $\phi$ gets scanned, the observed population of state |b> varies sinusoidally. Specifically, the population depends on this phase $\delta\phi$ as follows [1,2]:

$$P = \frac{1}{2}[1 - \cos(\delta\phi)] \,. \tag{34}$$

If there is no rotation, (i.e., $\delta\phi_0=0$), then the fringe minimum occurs at $\phi=0$. In the case of a non-zero rotation, this minimum is shifted to $\phi = -\delta\phi_0$. Measurement of this shift can therefore be used to determine the angular velocity from Eq. (33).

In order to establish a framework for interpreting the behavior of the CI, let us now consider a system where instead of doing a scan by applying a phase-shift only to the third beam (the last $\pi/2$ pulse), we apply a phase-shift partway through the middle beam (the $\pi$ pulse), which is of time length $\tau$ and space length $l$. This modified Borde-Chu Interferometer is shown in Fig. 3. These length parameters have the relationship $l = v_x\tau$, where $v_x$ is the velocity of the atom in the **x** direction. The phase-shift is applied starting from a distance $\delta l$ away from the center of the $\pi$ pulse. The center of the pulse corresponds to $\delta l = 0$, and the phase-shift $\phi$ is applied at all points in the beam to the right of $\delta l$. Thus, the $\pi$ pulse is effectively split into two beams, the first one of length $l/2+\delta l$ where there is no phase-shift applied, and the second of length $l/2-\delta l$ where the phase-shift $\phi$ is applied.

In deriving an approximate analytic expression for the complex amplitude $c_b=\langle b|\Psi\rangle$ of the state $|b\rangle$ after passing through such an interferometer, we use the Hamiltonian given by Eq. (28) and set $\Delta = 0$. This amounts to neglecting the Doppler shift and assuming that both lasers are equally detuned. We model the $\pi$ pulse as two separate beams of variable lengths. The first beam is of time length $\tau/2 + \delta\tau$, and the second is $\tau/2 - \delta\tau$. Letting $\tau_2 = \tau/2 - \delta\tau$ and noting that $\Omega_0\tau = \pi$, where $\Omega_0$ is the Raman Rabi frequency, $\omega_b$ is the free space propagation frequency, we can derive the expression for the amplitude of the excited state at the end of the $\pi/2$ pulse at the 3rd zone,

$$\begin{aligned} c_b = \frac{i}{2}\exp(-i\omega_b 2T)\Bigg[ & \sin\left(\frac{\Omega_0\tau_2}{2}\right)\cos\left(\frac{\Omega_0\tau_2}{2}\right)(\exp(-i\phi_{30})-1)(1-\exp(-i\phi)) \\ & +\cos^2(\frac{\Omega_0\tau_2}{2})\exp(-i\phi_{20})(\exp(-i\delta\phi_0 - i\phi)-1) \\ & +\sin^2(\frac{\Omega_0\tau_2}{2})\exp(-i\phi_{20})(\exp(-i\delta\phi_0)-\exp(-i\phi))\Bigg]. \end{aligned} \tag{35}$$

In the limit that $\tau_2 = 0$ or $\tau_2 = \tau$, and if the rotation velocity $\Omega$ is 0, the absolute value squared of this equation reduces to Eq. (34) for the fringes.

We are now in a position to compare the rotation sensitivity of the original BCI and that of this modified configuration where the phase is applied starting at some point in the $\pi$ pulse. First, we define the effective area $A_{eff}$ for the modified BCI as a proportionality constant between the calculated fringe shift $\Delta\phi$ and the rotation rate $\Omega$, in the following form:

$$\Delta\phi = -2m\Omega A_{eff}/\hbar . \tag{36}$$

This effective area may or may not be the same as the true area of the interferometer. In order to determine the value of $A_{eff}$, we require an expression for this fringe shift that results upon rotation for this new system. To derive this, we first take the absolute square of Eq. (35). Then take the derivative of the absolute square with respect to $\phi$ to find the minimum, and compare how far the minimum shifts as a function of rotation. After a tedious but straightforward calculation we get an exact expression for the phase shift $\Delta\phi$:

$$\Delta\phi = \tan^{-1}\frac{[\sin^4(\frac{\Omega_0\tau_2}{2}) - \cos^4(\frac{\Omega_0\tau_2}{2})]\sin(\delta\phi_0)}{\sin^4(\frac{\Omega_0\tau_2}{2})\cos(\delta\phi_0) - \frac{1}{2}\sin^2(\Omega_0\tau_2)\cos(2\delta\phi_0) + \cos^4(\frac{\Omega_0\tau_2}{2})\cos(\delta\phi_0)}. \quad (37)$$

In the limit where $\delta\phi_o$ is small, Eq. (37) reduces to:

$$\Delta\phi = -\tan^{-1}\frac{\delta\phi_o}{\sin(\Omega_0\delta\tau)}. \quad (38)$$

As expected, we see that in the limits of $\delta\tau = \pm\,\tau/2$, the fringe shift approaches the value of $\mp\delta\phi_o$ in Eq. (33). Therefore, in these limits, $A_{eff}=A_o$. As $|\delta\tau|$ becomes smaller than $\tau/2$, the magnitude of the fringe shift actually becomes bigger than $\delta\phi_0$. However, this does not represent any improvement in our ability to measure the rate of rotation, due to the fact that the fringe amplitudes become smaller in the same time. This can be appreciated immediately by noting that in the limit of $\delta\tau \to 0$, the fringe shift approaches $\pi/2$, independent of the rate of rotation, but the fringe amplitude approaches zero.

In order to interpret this result quantitatively, it is instructive to define a *minimum measurable rotation rate*: $\Omega_{mm}$. By rearranging Eq. (33), we see that $\Omega_{mm}$ depends on the minimum measurable fringe shift $\Delta\phi_{mm}$:

$$\Omega_{mm} = \frac{\hbar}{2m}\frac{\Delta\phi_{mm}}{A_{eff}}. \quad (39)$$

The rotational phase shift is determined from the horizontal shift of the phase scan. The minimum measurable fringe shift has to be greater than the amplitude of the noise on the phase scan. Therefore, if the amplitude of the phase scan is $\alpha$ ($\leq 1$), and the signal amplitude is $S \equiv \alpha S_o$ (where $S_o$ is the maximum signal, determined by the number of atoms, the detection efficiency, and the integration time), with the amplitude of the noise being N, $\Delta\phi_{mm}$ is given by

$$\Delta\phi_{mm} = \frac{\pi}{S/N}. \quad (40)$$

Assuming shot-noise limited detection, the signal to noise ratio is $\sqrt{S}$, so that the minimum measurable phase shift is $\pi/\sqrt{S}$, and the minimum measurable rotation rate is

$$\Omega_{mm} = \frac{\hbar}{2m} \frac{\pi}{A_{eff}\sqrt{\alpha S_o}} . \tag{41}$$

Under ideal conditions, the amplitude of the phase scan for the original BCI is 1, from equation (34). Therefore, the minimum measurable rotation rate for a BCI where the phase is applied only in the last $\pi/2$ pulse is

$$\Omega_{mm(BCI)} = \frac{h}{4m} \frac{1}{A_o\sqrt{S_o}} , \tag{42}$$

where $A_o$ is the true area for the BCI. Define the quality factor Q as the ratio between the minimum measurable rotation rates of the BCI with phase applied in the last pulse and with the phase applied in the middle pulse $(MBCI)'$:

$$Q = \frac{\Omega_{mm(BCI)}}{\Omega_{mm(MBCI)'}} = \frac{1/A_o}{1/[A_{eff}\sqrt{\alpha}]} . \tag{43}$$

If we define the ratio η between the areas as $\eta = A_{eff}/A_0$ , Q becomes

$$Q = |\eta|\sqrt{\alpha} . \tag{44}$$

Thus, if Q > 1, the minimum measurable rotation rate of the modified BCI system is smaller than that of the original BCI. This provides us with a framework for comparison of different kinds of interferometer systems, with respect to their rotation sensitivity. We can now directly compare the BCI system that has the phase applied in the middle pulse with the original BCI by plotting the quality factor Q vs. $\delta\tau$. For this we need the signal amplitude as a function of $\delta l$, which is easily calculated from Eq. (35):

$$\alpha = \cos^2(\Omega_0\tau_2) = \sin^2(\Omega_0\delta\tau). \tag{45}$$

From Eqs. (33), (36) and (38), we get

$$\eta = \frac{1}{\delta\phi_o}\tan^{-1}\frac{\delta\phi_o}{\sin(\Omega_0\delta\tau)} . \tag{46}$$

Therefore, the quality factor is

$$Q = \frac{\sin(\Omega_0\delta\tau)}{\delta\phi_0}\tan^{-1}\left(\frac{\delta\phi_0}{\sin(\Omega_0\delta\tau)}\right) . \tag{47}$$

Figure 4 shows a typical plot of η, for $\Omega_0 = 2\pi(7 \times 10^4)$ $s^{-1}$, $L = 3 \times 10^{-3}$ m, and $k = 8.055 \times 10^6$ $m^{-1}$ (corresponding to the D2 transition in Rb), and $\delta\phi_o$=0.1. Note that the effective area approaches $A_0$ as $\delta\tau/\tau \rightarrow \pm 0.5$, as expected. As $\delta\tau$ goes to 0, the effective area approaches $\pi A_0 / 2\delta\phi_0 = 5\pi A_0$. Note that this large effective area does not imply that the interferometer actually encloses such a large area; rather, it is a convenient way to quantify the fact that the fringe shift for a given rotation is larger. However, as indicated above, this does not imply any increase in the ability to detect the rotation rate, since the signal amplitude drops to 0 in this limit. Thus, the quality factor decreases and becomes 0 very rapidly. These are illustrated in figure 5, which shows α and Q for the same set of parameters. Note that if the phase is applied away from $\delta\tau$= 0, then the quality factor remains very close to unity.

Now we proceed to investigate the behavior of the continuous interferometer with respect to similar variables. Our setup for the CI differs in particular from the BCI in that the atom traverses only a single laser beam having a Gaussian electric field profile in the transverse direction, as illustrated in Figs. 6 and 7. As the atom passes through the beam, the wavepackets for the |a> and |b> states take different trajectories depending on the width of the beam and the effective Rabi frequency $\Omega_0$. In order to do a phase scan in this system, we apply a phase-shift to this laser pulse starting from some position $\delta l$ measured from the center of the pulse and extending in the direction of propagation of the atom, as shown in Fig. 7. Such a scan can be realized by placing a glass plate in the path of the beam, inserted only partially into the transverse profile of the laser beams, and rotating it in the vertical direction. Any potential problem of diffraction can be eliminated by ensuring that the plat is placed close to the atomic beam, or by using imaging optics reverse the diffraction. We see that this configuration is analogous to the BCI system analyzed previously where the phase-scan is applied starting from somewhere in the second beam. If this interferometer is made to rotate, there will again be a rotational fringe shift. We expect that there will be a variation of the effective area and the signal amplitude α with $\delta l$, and that this variation will be similar to that of the modified BCI.

In order to calcualte the fringe shift for the CI, we use the formalism developed above, which are summarized in Eqs. (30)-(32). We imagine the laser profile being sliced up into infinitesimal intervals Δx in the transverse direction. Each one of these slices is rotating with angular velocity Ω, but will have a different deviation in the y direction depending on how far away it is from the axis of rotation. This will lead to the atom seeing a different phase shift at every point x in the laser profile. In our simulations, we placed the axis of rotation at the point A in the diagram.

The phase shift for this interferometer is also linear for infinitesimal rotations. Thus, an effective area for this interferometer can be defined as in Eq. (36). We choose to simulate a system with the following parameters, $\Omega_0 = 2\pi\ (7\times10^4)$ and $L = 3 \times 10^{-3}$ m, such that $\Omega_0 T = 3.3$. The atom is a Gaussian wavepacket with a 1/e spread of 1/k, where $k = 8.0556 \times 10^6$ $m^{-1}$, corresponding to the wavelength of the laser, 780 nm. So the 1/e spread (half-width) of the atomic wave-packet is roughly 0.13 μm. The wavepacket centroid trajectories in the CI are shown in Figs. 8 through 11 for different applied phase-shifts φ = 0, π/2, π, and 3π/2, respectively. For these trajectories, the phase-shift is applied starting at $\delta l/l$ = 12/25. In these figures, the trajectories may appear to be completely different from one another; however, note that the atomic wave-packets are highly overlapped, since the 1/e half-width is about 0.13 μm. The trajectories are plotted with no rotation in the system. If the system is rotating there will be slight deviations in the trajectories, which lead to the rotational fringe shifts. Simulations were performed to determine these fringe shifts as a function of the point of application of the phase-scan. This in turn was used to determine the effective area of the interferometer. The signal amplitude was also determined as a function of $\delta l/l$. The variations of the signal amplitude $\alpha$, the effective area $A_{eff}$, and the quality factor Q as a function $\delta l/l$ are plotted in Figs. 12, 13, and 14, respectively. The maximum fringe contrast for our system is 0.955 and occurs at $\delta l/l = \pm 0.48$. The phase scan showing this result is in Fig. 15.

In order to compare this rotation sensitivity with that of a BCI, we now need to know the area of the BCI that would correspond to the parameters of our system, i.e., the CI. To make this correspondence, we note that most of the interaction in the Gaussian laser profile occurs within one standard deviation of the peak of the profile. Thus, it is reasonable to define an equivalent BCI with a zone-separation length of $L = 3 \times 10^{-3}$ m (so that the three-zone length is 2L), which is the 1/e length of the Gaussian profile. The area of a BCI is given by the following formula

$$A_0 = L^2 \frac{2\hbar k/m}{v_x}. \tag{48}$$

For $L = 3 \times 10^{-3}$ m, we get $A_0 = 2.7 \times 10^{-10}$ m$^2$.

With this value of $A_0$ we can go through similar steps as for the BCI and work out the variation of the quality factor Q as a function of $\delta l/l$. The values of the relative effective area $\eta$, the fringe amplitude $\alpha$ and the quality factor Q versus $\delta l/l$ are plotted in Figs. 16 and 17. The quality factor for the CI has a shape very similar to the BCI. However, in contrast to the BCI, the effective area varies smoothly through 0, which affects the variation of the quality factor as well. The signal amplitude also never reaches 0, as it does in the BCI. The quality factor is approximately one for $|\delta l/l| > 0.25$, which means that if the phase-scan is applied starting in this range of values, *our interferometer will provide the same rotation sensitivity as a BCI of the same size.*

Note that the above results for the rotational fringr shifts and effective areas do not depend on whether the shift is measured with the applied phase at the minimum or the maximum of the phase scan. This is true not only for the BCI but also for our continuous interferometer. In the case of the CI, however, this result has a rather surprising implication about the relationship between the effective area and the trajectories of the |a> and |b> wavepackets. As shown in Figs. 8 through 12, the trajectories of the wavepackets are different for different values of the applied phase shifts. Thus, if the effective area were dependent upon the wavepacket trajectories, it would vary with changes in the applied phase. The fact that the effective area does not change with regard to where in the phase scan the applied phase is located demonstrates that the effective area $A_{eff}$ of the CI is independent of the wavepacket trajectories.

## 4. Remarks

The CI is operationally simpler because it uses only one zone. In practice, this means that there is no need to ensure the precise parallelism of the three zones, as needed for the BCI. As such, the CI may be preferable to the BCI, given that the rotational sensitivity of the CI can be same as that of the BCI. One potential concern is that while the BCI can accommodate an effective length (separation between the first and the third zones is 2L) as long as several meters, such a long interaction length for the CI would be impractical. On the other hand, an interferometer that is several meters long is unsuited for practical usage such as inertial navigation. Therefore, it is likely that a practical version of the BCI would be much shorter (several cm's) in length. Of course, such a short length would reduce the rotational sensitivity drastically. This problem can potentially be overcome by using a slowed atomic beam (e.g., from a magneto-optic trap or Bose-condensate), so that the transverse spread of the split beams would be much larger, thereby compensating for the reduction in the longitudinal propagation distance. Under such a scenario, the CI would be simpler than the BCI, while yielding the same degree of rotational sensitivity.

## 5. Acknowledgement

This work was supported by DARPA Grant No. F30602-01-2-0546 under the QUIST program, ARO Grant No. DAAD19-001-0177 under the MURI program, and NRO Grant No. NRO-000-00-C-0158.

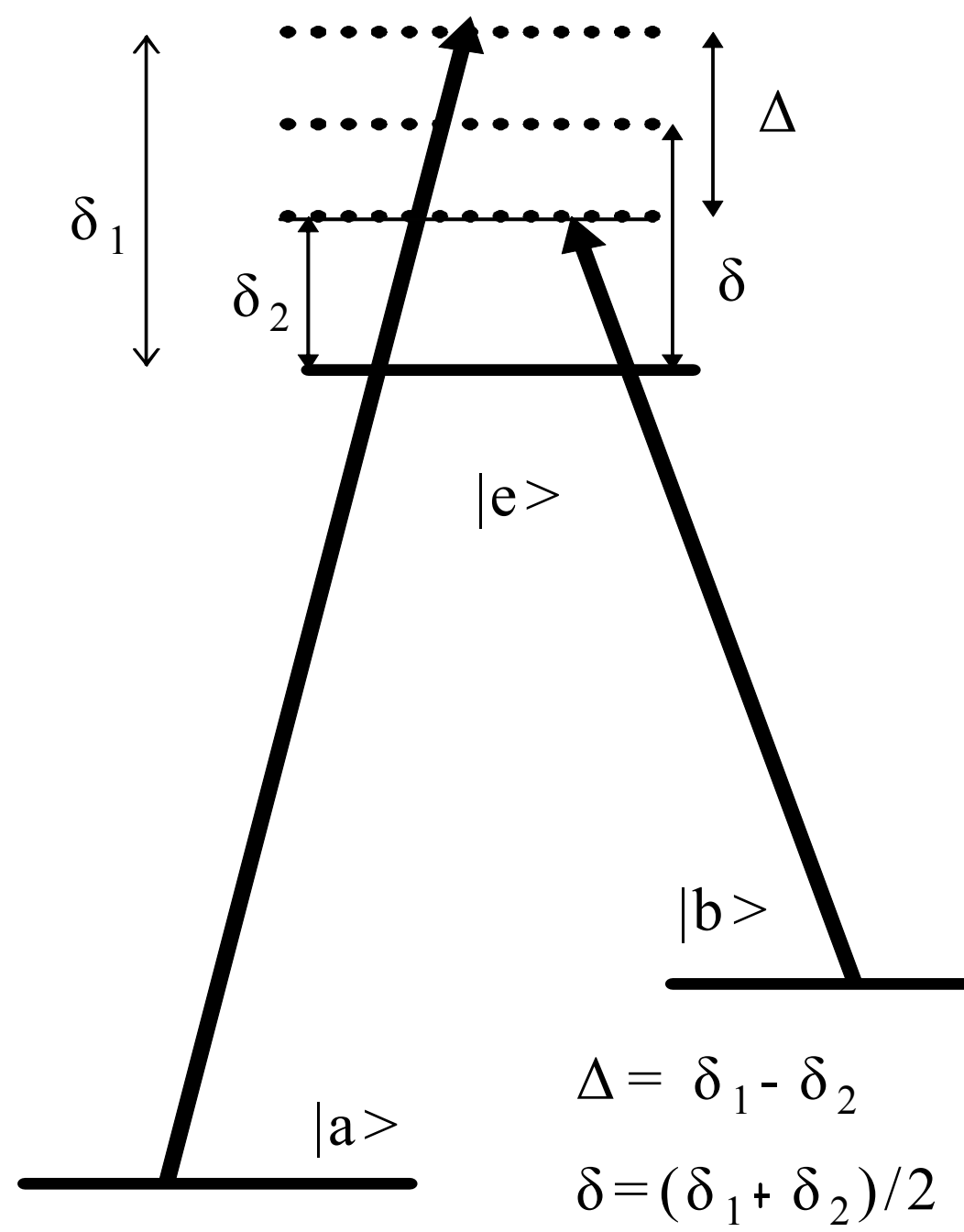

FIG. 1. A schematic picture of a three level system where $\delta$ is the common detuning and $\Delta$ is the difference detuning.

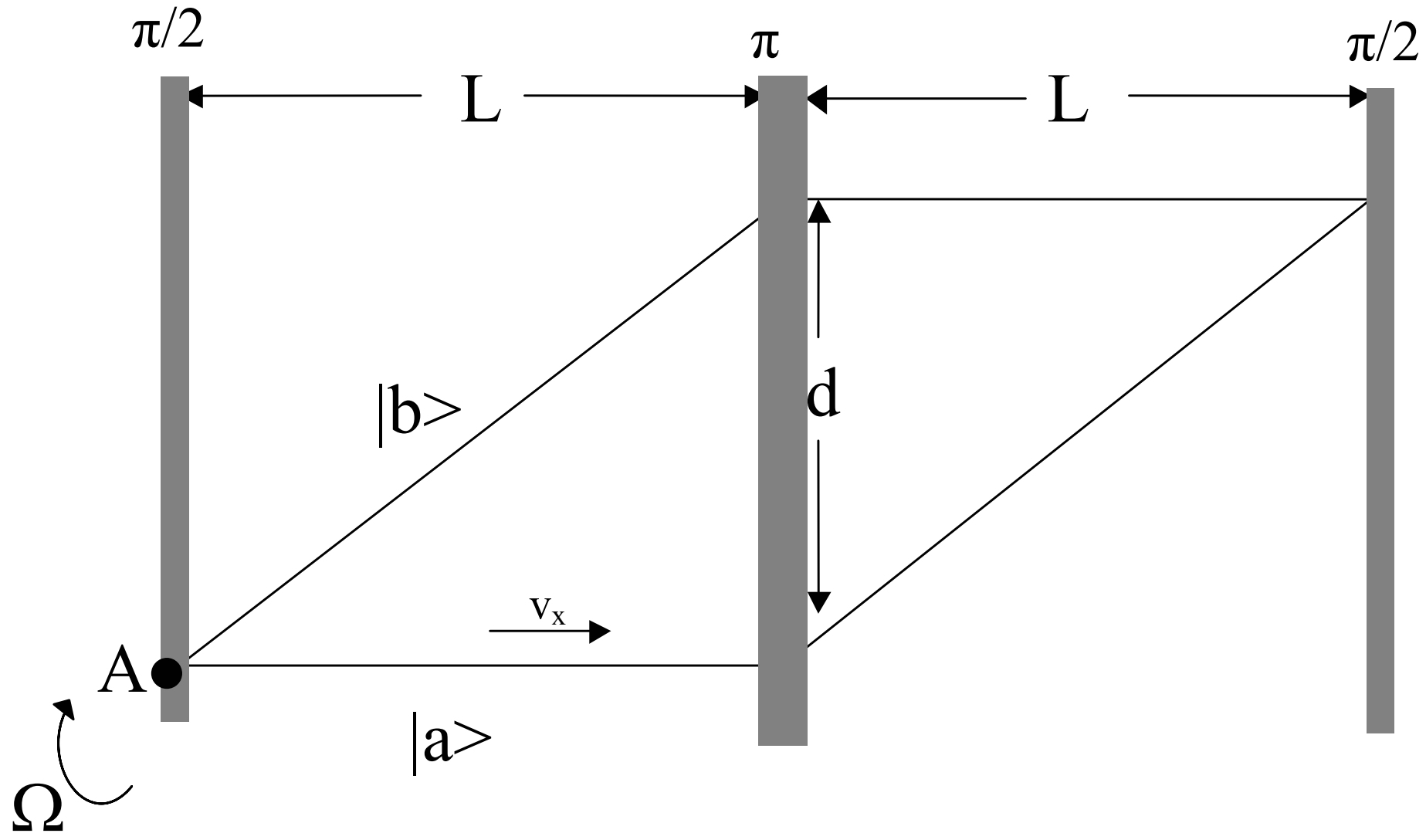

FIG. 2. Schematic of the setup for the Borde-Chu Interferometer. The atom moves at some speed $v_x$, with the two atomic states deflected along different paths. The entire system is rotating around point A at some rotational velocity Ω.

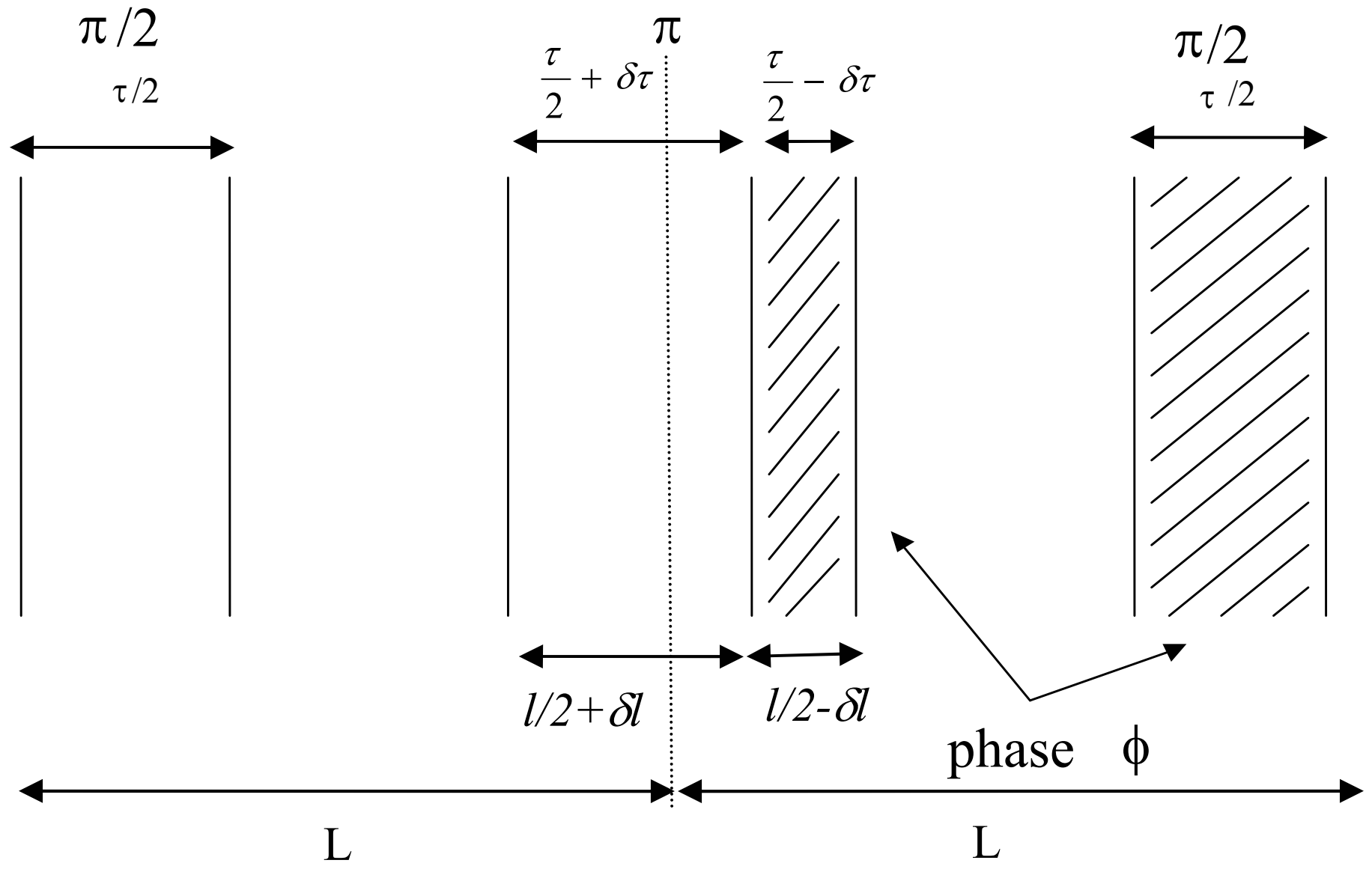

FIG. 3. Modified Borde-Chu Interferometer where the phase is applied partway through the center $\pi$ pulse and through the last $\pi/2$ pulse. The phase is applied starting at $\delta l$ from the middle of the $\pi$ pulse.

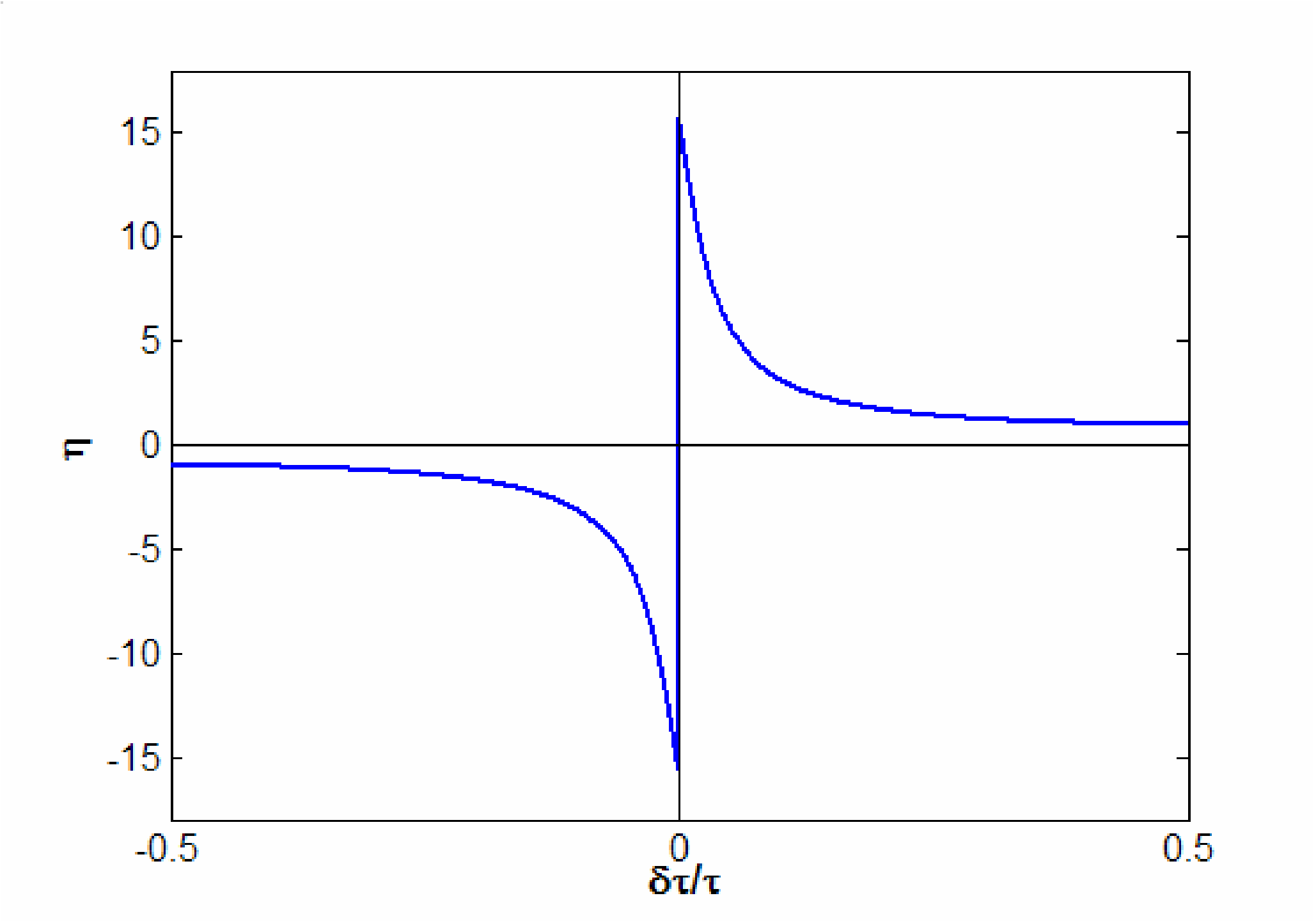

FIG. 4. A typical plot of η, for $\Omega_0 = 2\pi(7 \times 10^4)$ s$^{-1}$, L = 3 × 10$^{-3}$ m, and k = 8.055 × 10$^6$ m$^{-1}$ (corresponding to the D2 transition in Rb), and $\delta\phi_o$=0.1. Note that the effective area approaches $A_0$ as $\delta\tau/\tau \rightarrow \pm 0.5$, as expected. As $\delta\tau$ goes to 0, the effective area approaches $\pi A_0 / 2\delta\phi_0 = 5\pi A_0$.

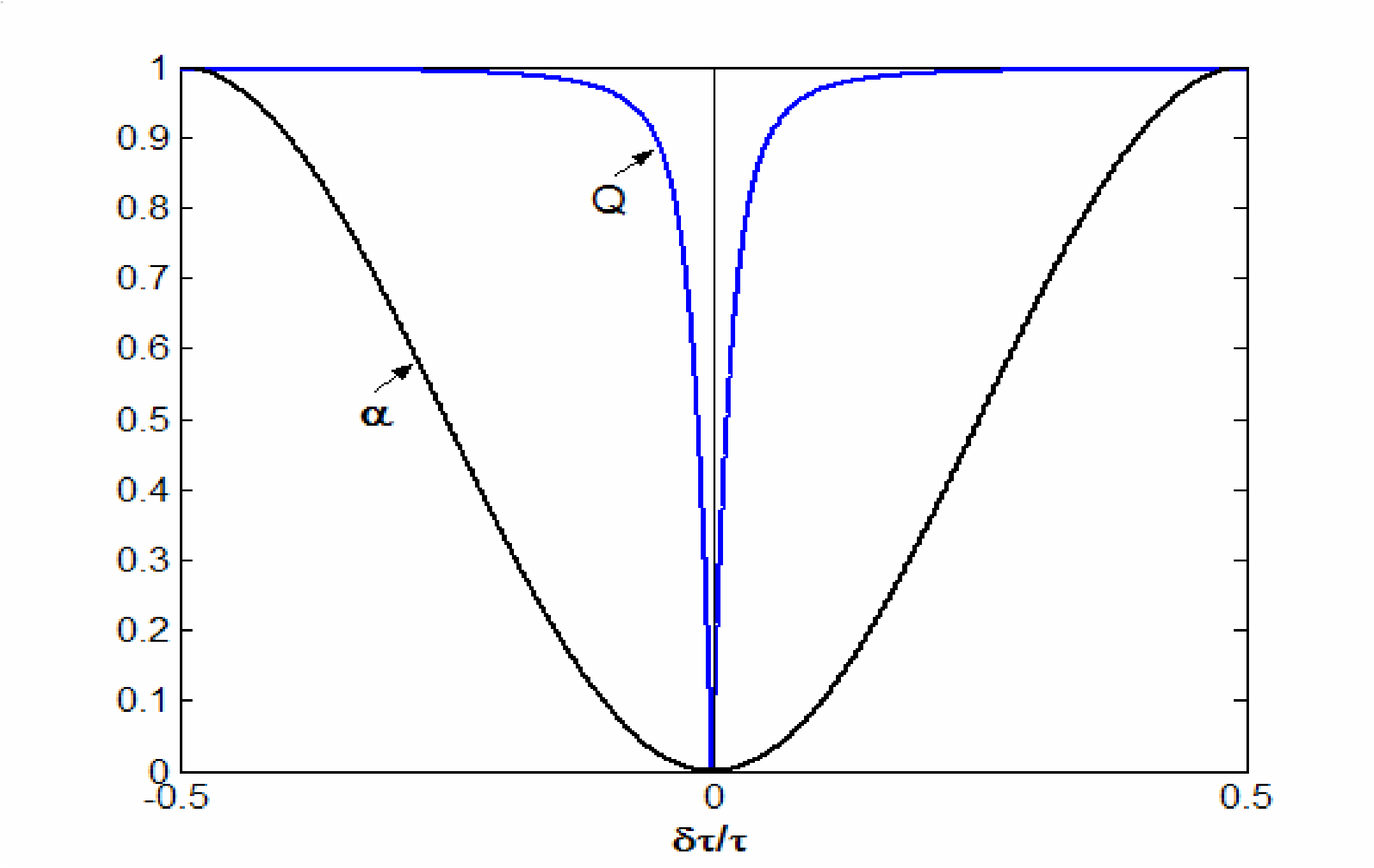

FIG. 5. Typical plots of $\alpha$ and Q versus $\delta\tau/\tau$, for $\Omega_0 = 2\pi(7 \times 10^4)$ $s^{-1}$, $L = 3 \times 10^{-3}$ m, and $k = 8.055 \times 10^6$ $m^{-1}$ (corresponding to the D2 transition in Rb), and $\delta\phi_o$=0.1. Note that the quality factor never exceeds unity, and drops to zero as $\delta\tau$ goes to 0.

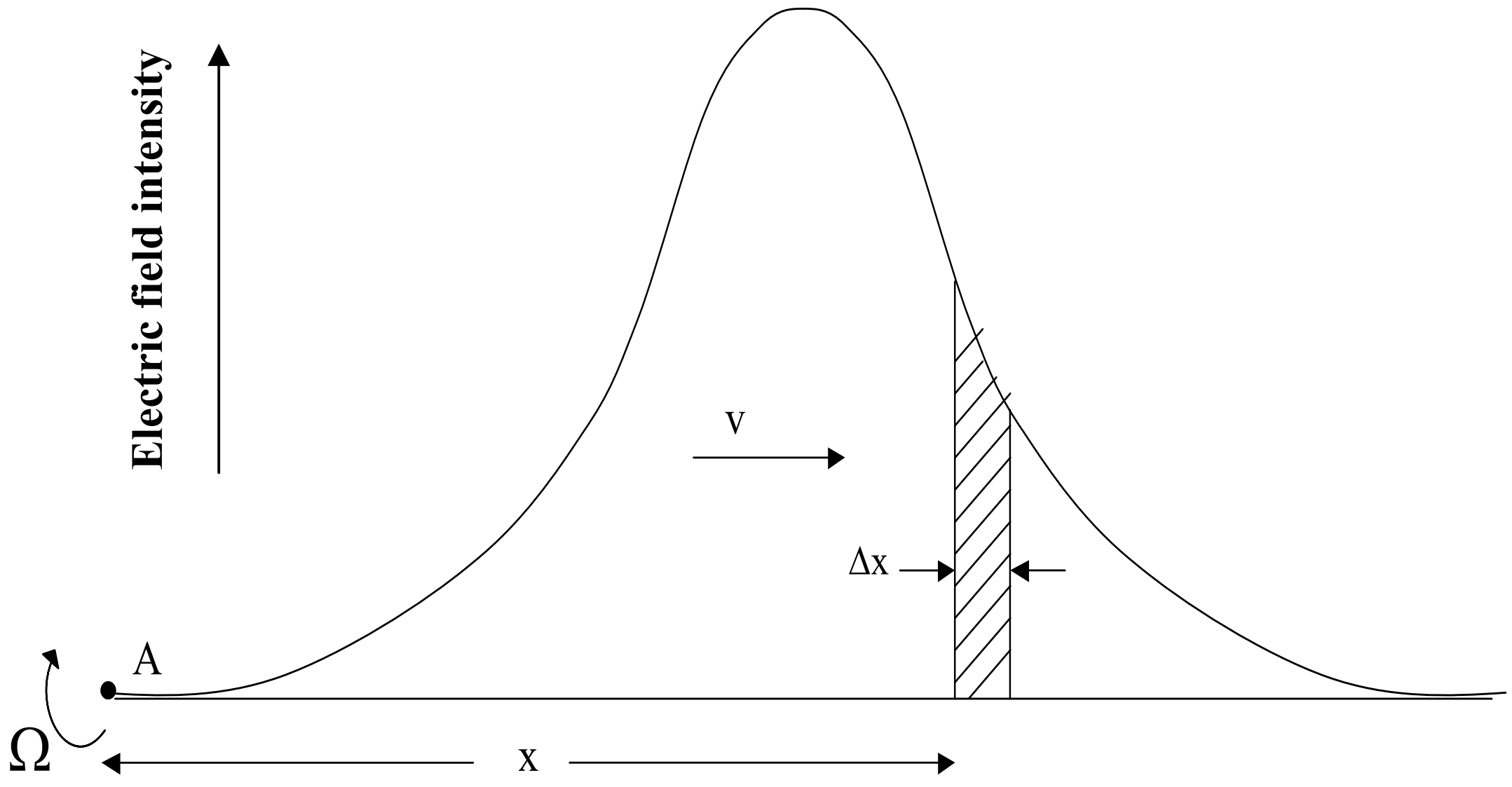

FIG. 6. Electric field intensity profile of counter-propagating laser beams for the Continuous Interferometer. The field intensity of the laser beam varies in the x direction as a Gaussian. The entire setup is rotating around point A, with the atom moving at speed v in the x direction.

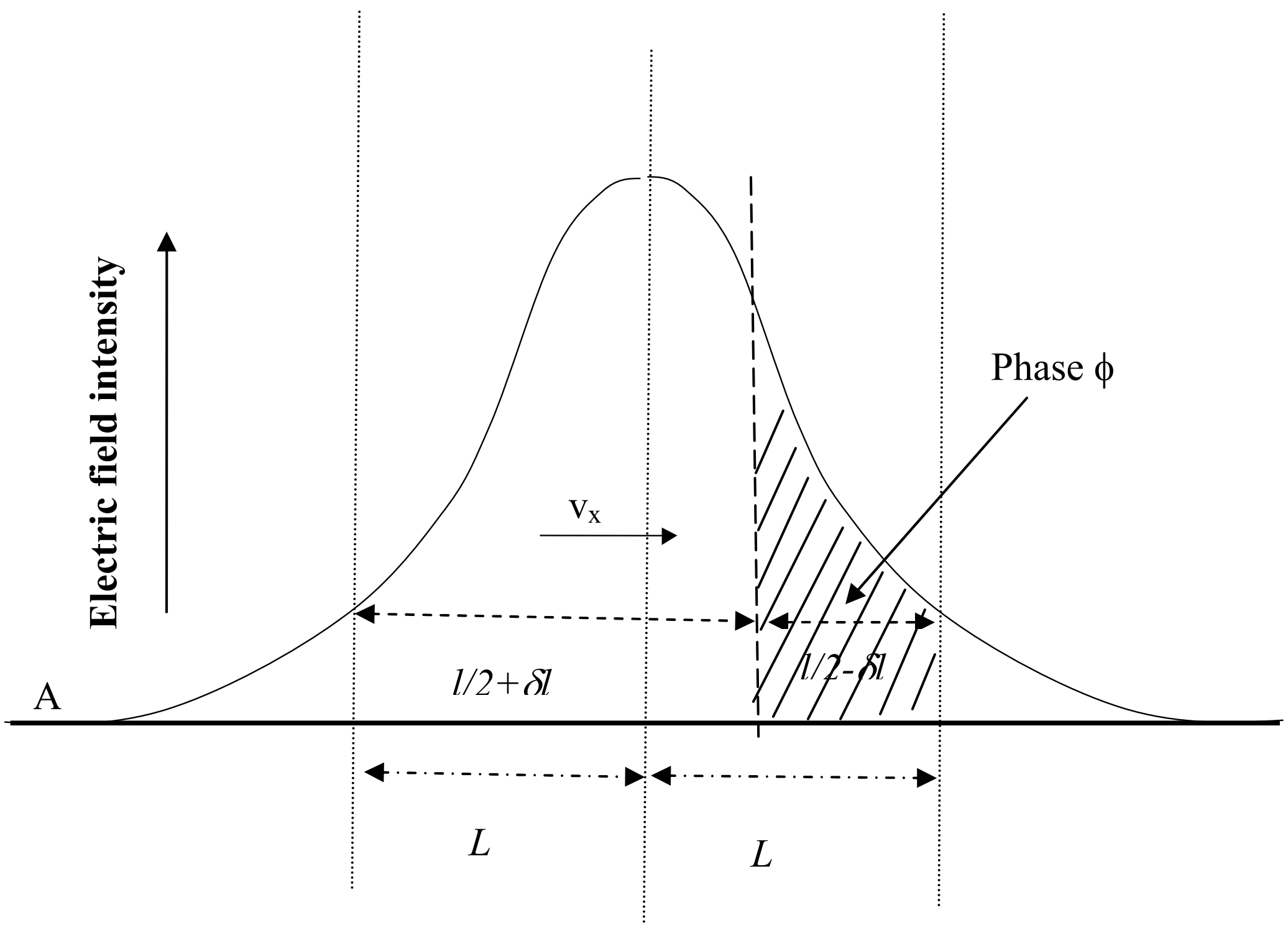

FIG. 7. Phase-shift application in the continuous interferometer (CI). The phase is applied after a distance $\delta l$ from the middle of the beam as indicated by the shaded region. The velocity of the atom along the x direction $v_x$ relates L and T: $L = v_x T$.

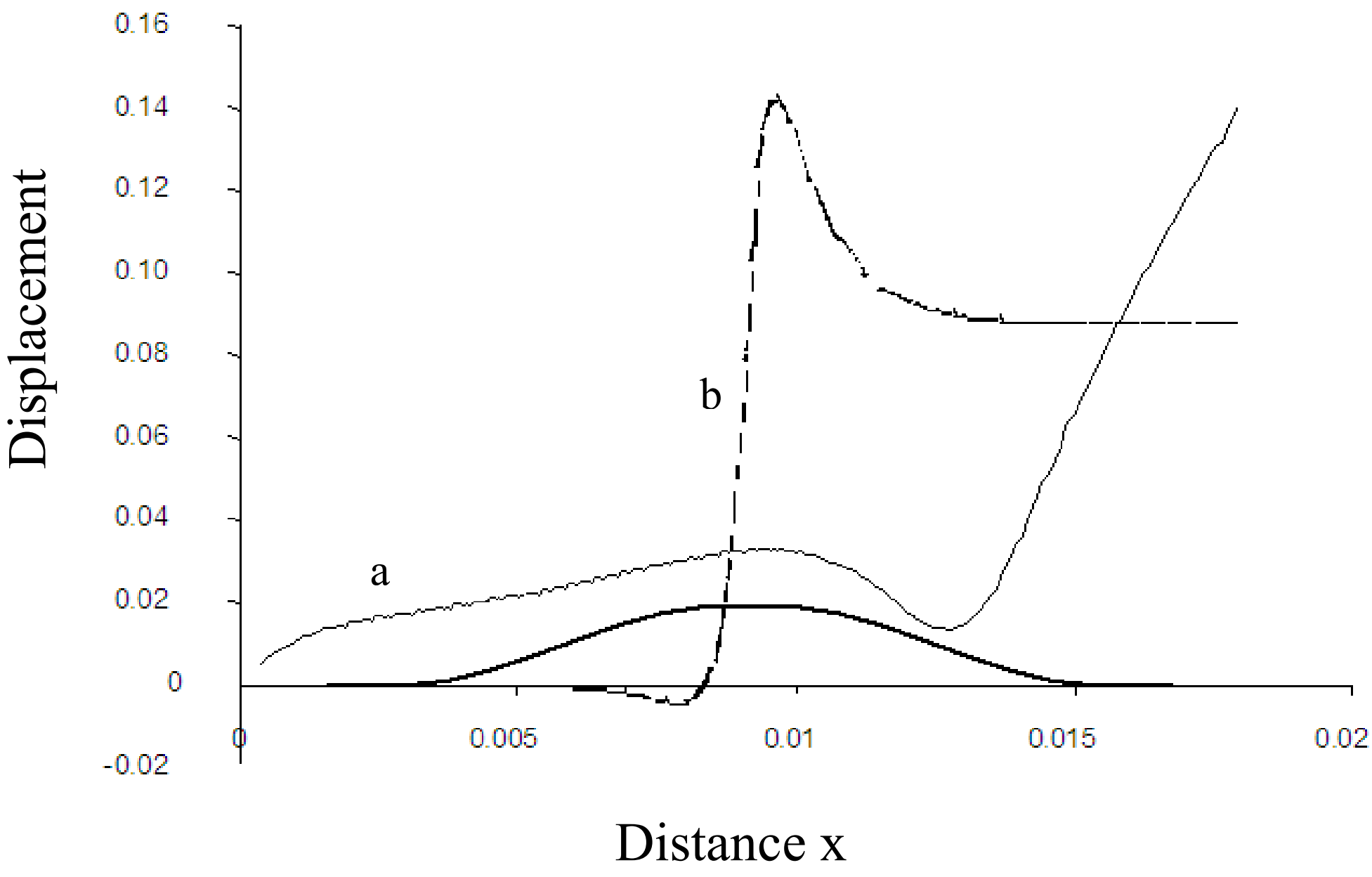

FIG. 8. Deflection (μm) of the centroids of the |a> and |b> state wavepackets vs. distance (m) into the laser beam. The |a> state wave packet trajectory is given by the solid line, the |b> state trajectory is the dashed line. The thick solid line is the Gaussian profile of the laser field intensity (in arbitrary unit along the intensity axis). No phase shift is applied in the laser beam. Although it may appear that the states are taking completely different trajectories, the wavepackets are in fact overlapping significantly since the 1/e length for the Gaussian position wavepackets is 0.13 μm. Simulated with parameters $\Omega_0 = 2\pi(7 \times 10^4)$, $L = 3 \times 10^{-3}$ m., $\Omega_0 T = 3.3$.

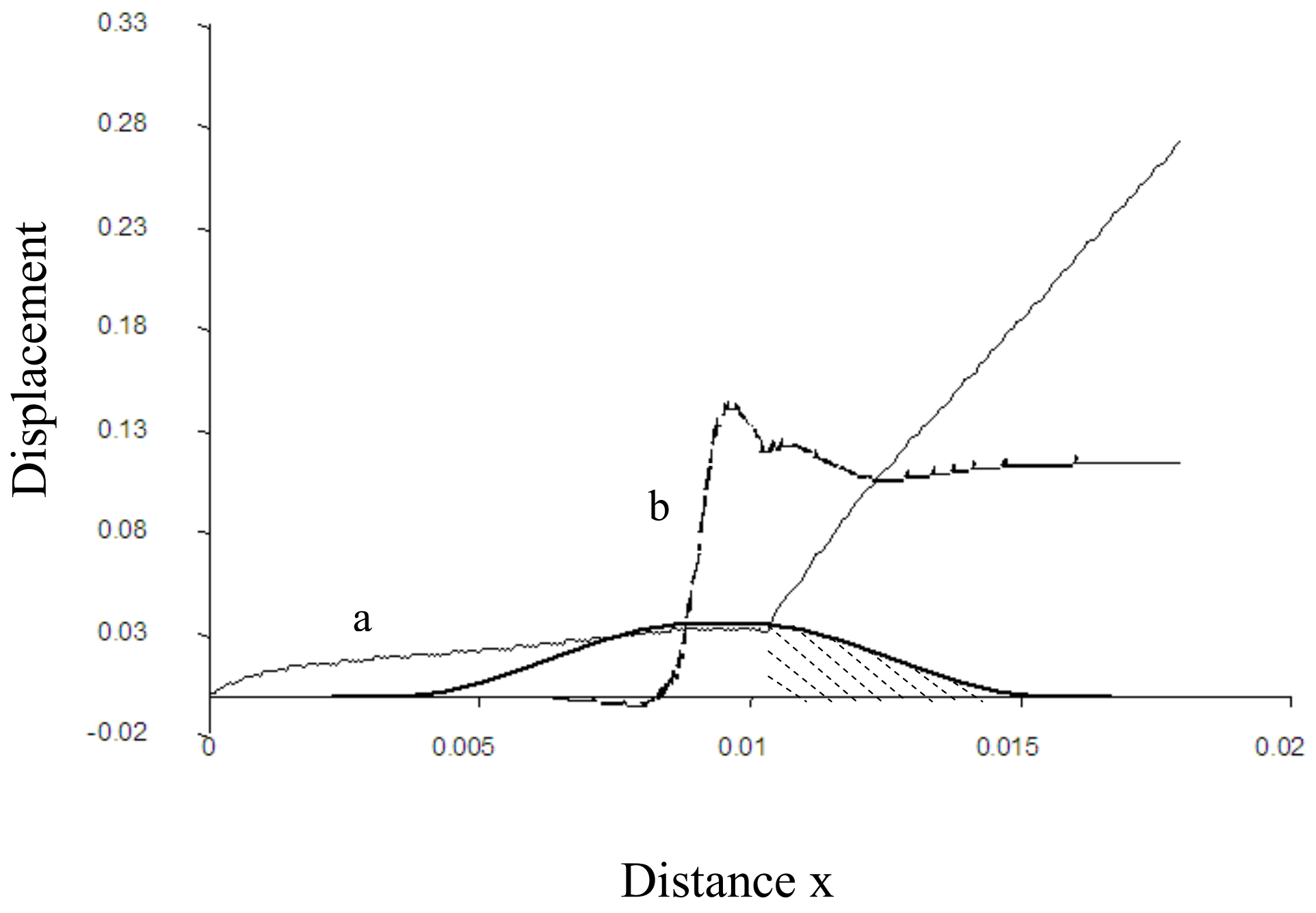

FIG. 9. Deflection (μm) of the centroids of the |a> and |b> state wavepackets vs. distance (m) into the laser beam. The |a> state wavepacket trajectory is given by the solid line, the |b> state trajectory is the dashed line . The thick solid line is the Gaussian profile of the laser field intensity (in arbitrary unit along the intensity axis), and a phase of $\pi/2$ is applied from $\delta l/l = 12/25$, where the profile is shaded. Simulated with parameters $\Omega_0 = 2\pi(7 \times 10^4)$, $L = 3 \times 10^{-3}$ m., $\Omega_0 T = 3.3$.

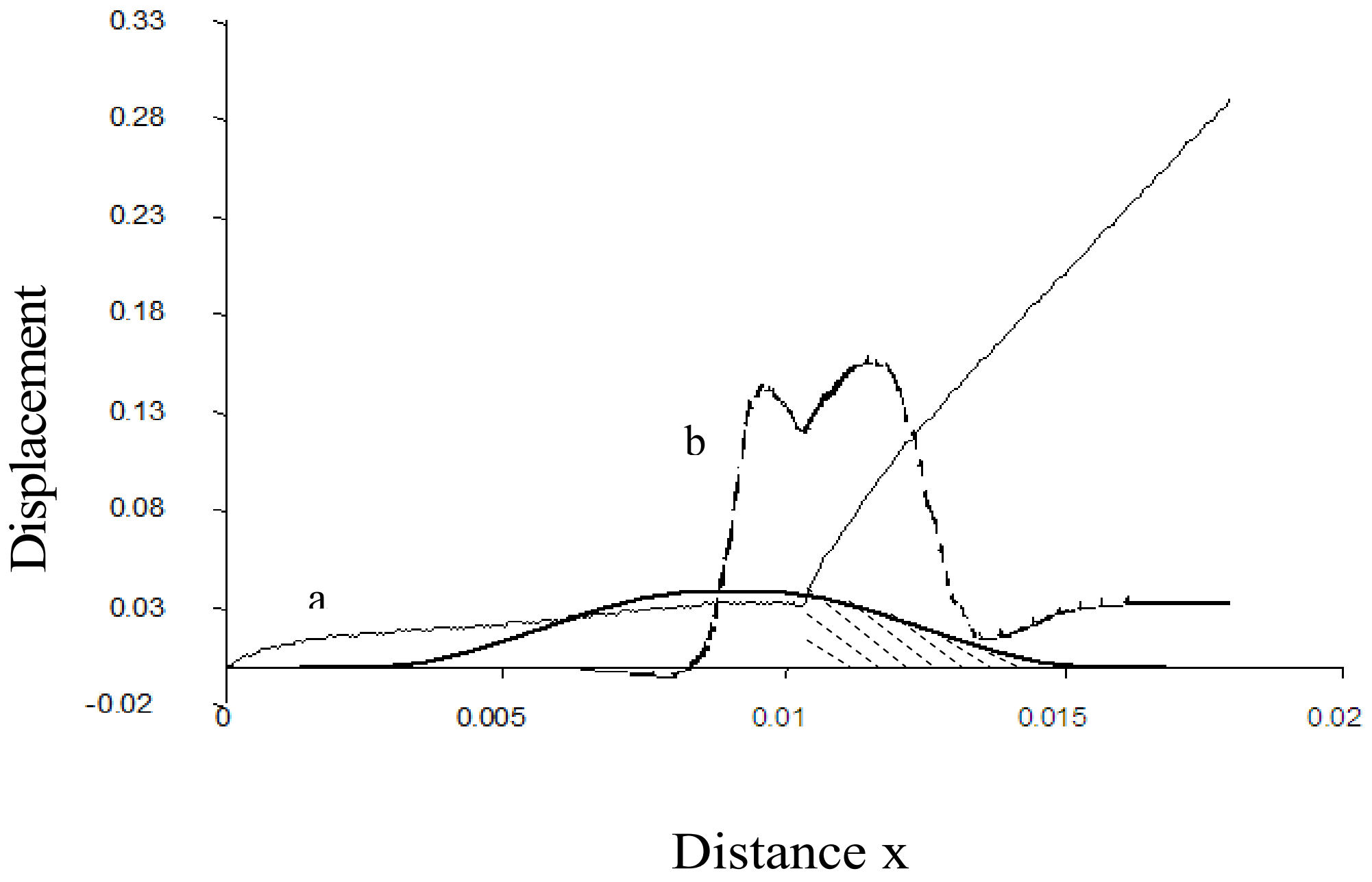

FIG. 10. Deflection (μm) of the centroids of the |a> and |b> state wavepackets vs. distance (m) into the laser beam. The |a> state wavepacket trajectory is given by the solid line, the |b> state trajectory is the dashed line. The thick solid line is the Gaussian profile of the laser field intensity (in arbitrary unit along the intensity axis), and a phase of $\pi$ is applied from $\delta l/l$ = 12/25, where the profile is shaded. Simulated with parameters $\Omega_0 = 2\pi(7 \times 10^4)$, $L = 3 \times 10^{-3}$ m., $\Omega_0 T = 3.3$.

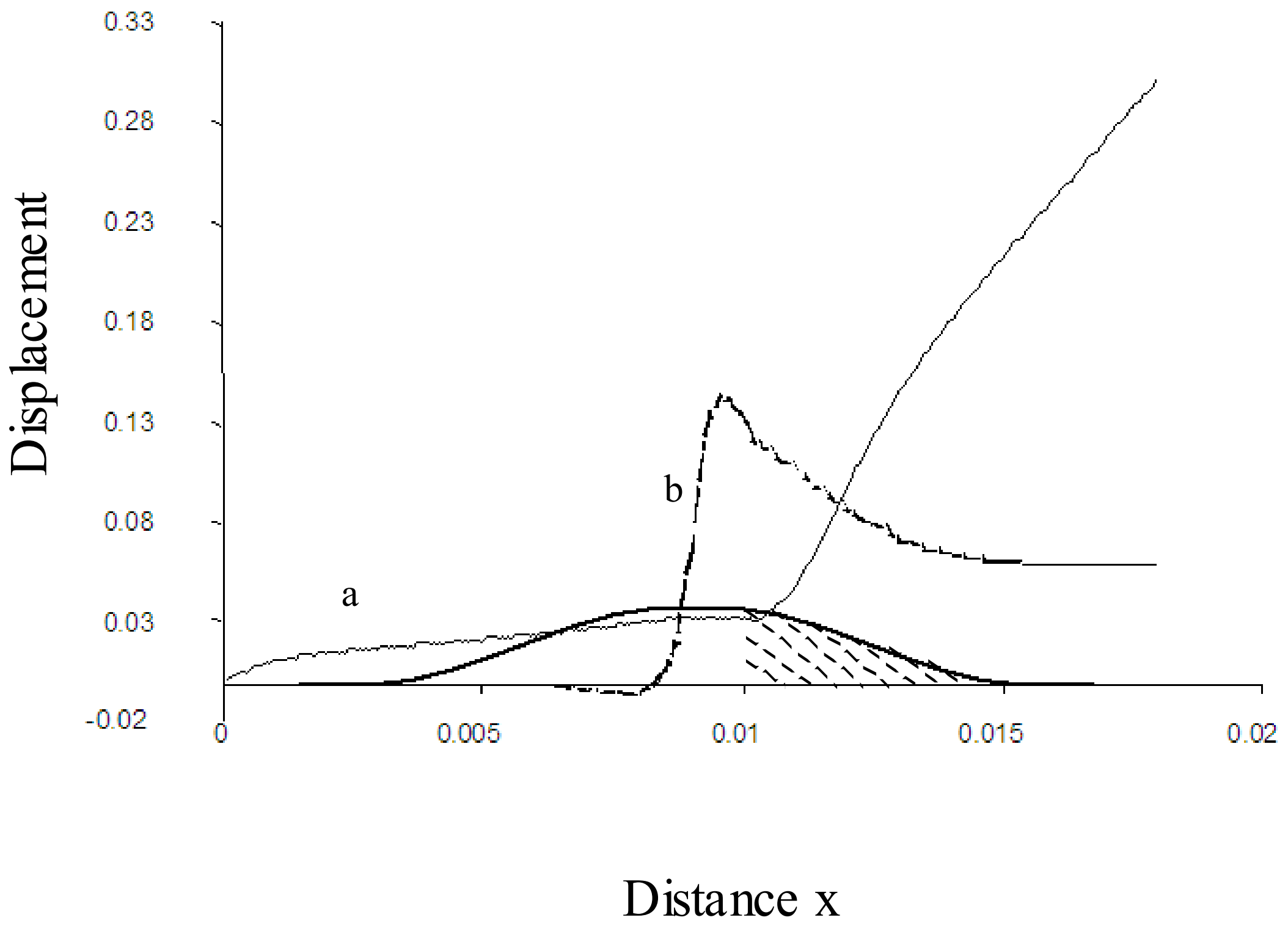

FIG. 11. Deflection (μm) of the centroids of the |a> and |b> state wavepackets vs. distance (m) into the laser beam. The |a> state wave packet trajectory is given by the solid line, the |b> state trajectory is the dashed line. The thick solid line is the Gaussian profile of the laser field intensity (in arbitrary unit along the intensity axis), and a phase of $3\pi/2$ is applied from $\delta l/l$ = 12/25, where the profile is shaded. Simulated with parameters $\Omega_0 = 2\pi(7 \times 10^4)$, $L = 3 \times 10^{-3}$ m., $\Omega_0 T = 3.3$.

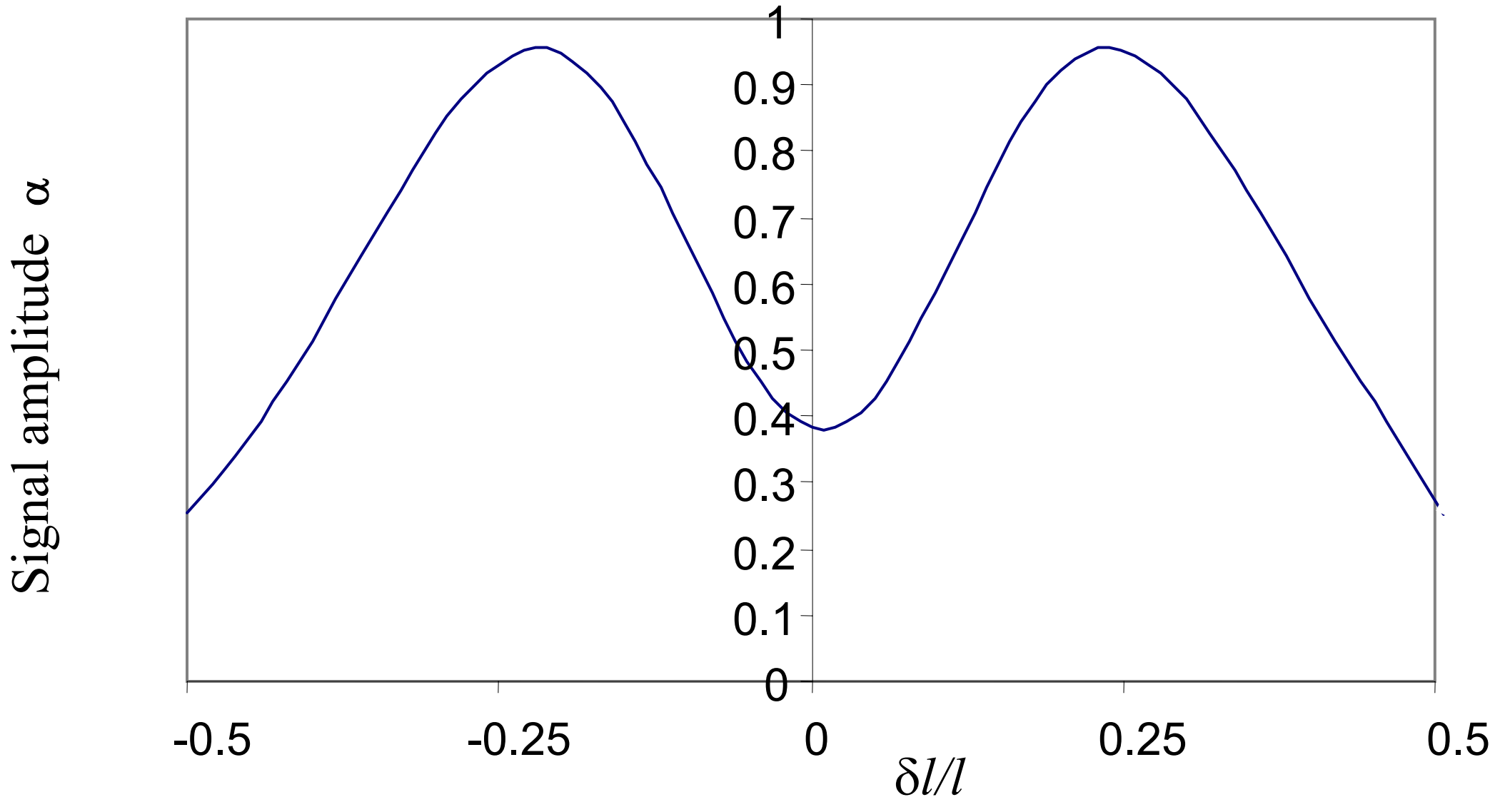

FIG. 12. Signal amplitude $\alpha$ vs. $\delta l/l$ for the Continuous Interferometer, simulated with parameters $\Omega_0 = 2\pi(7 \times 10^4)$, L = $3 \times 10^{-3}$ m., $\Omega_0 T = 3.3$.

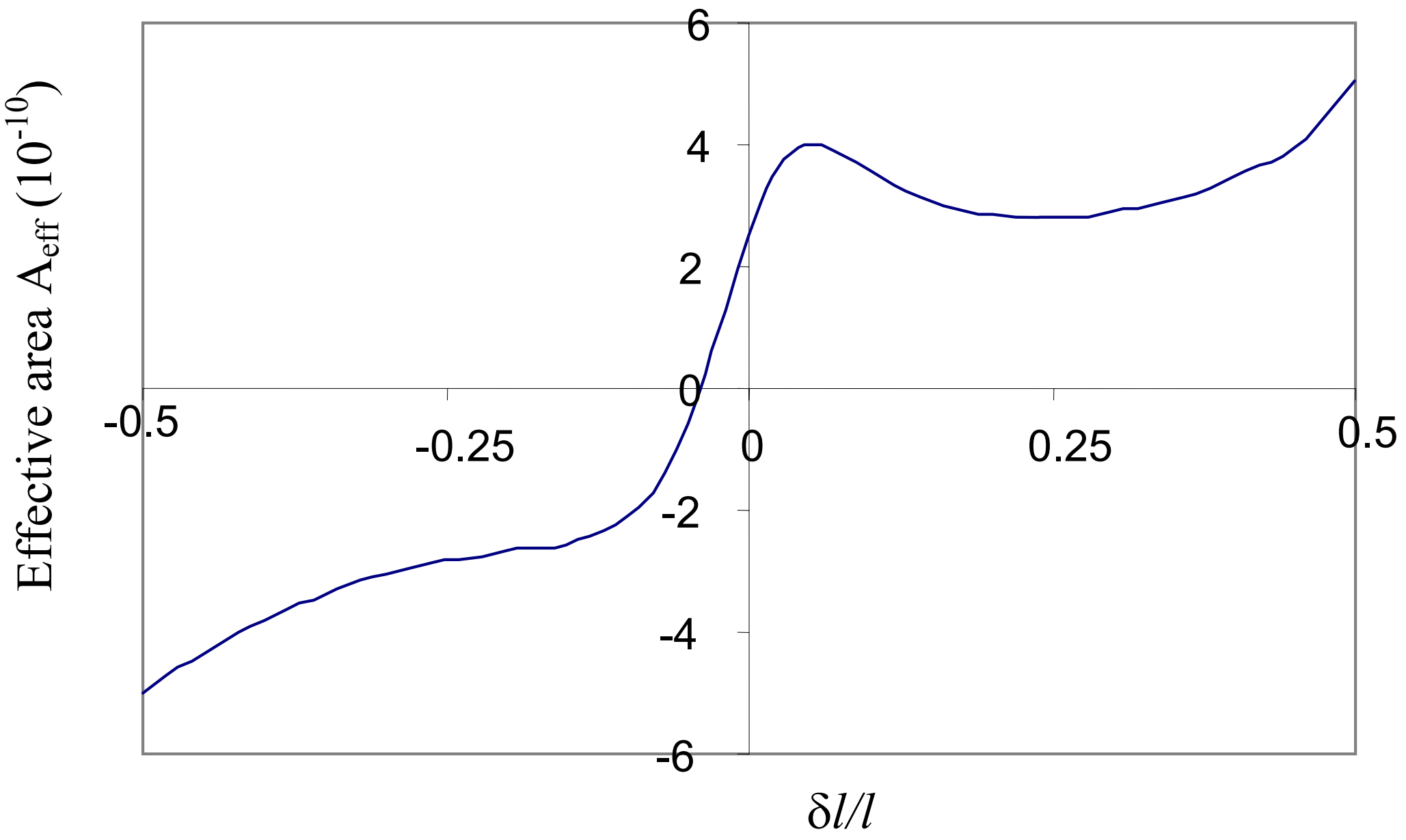

FIG. 13. Effective area ($m^2$) vs. $\delta l/l$ for Continuous Interferometer, simulated with parameters $\Omega_0 = 2\pi(7 \times 10^4)$, $L = 3 \times 10^{-3}$ m., $\Omega_0 T = 3.3$. The area for a comparable BCI is $2.7\times 10^{-10}$.

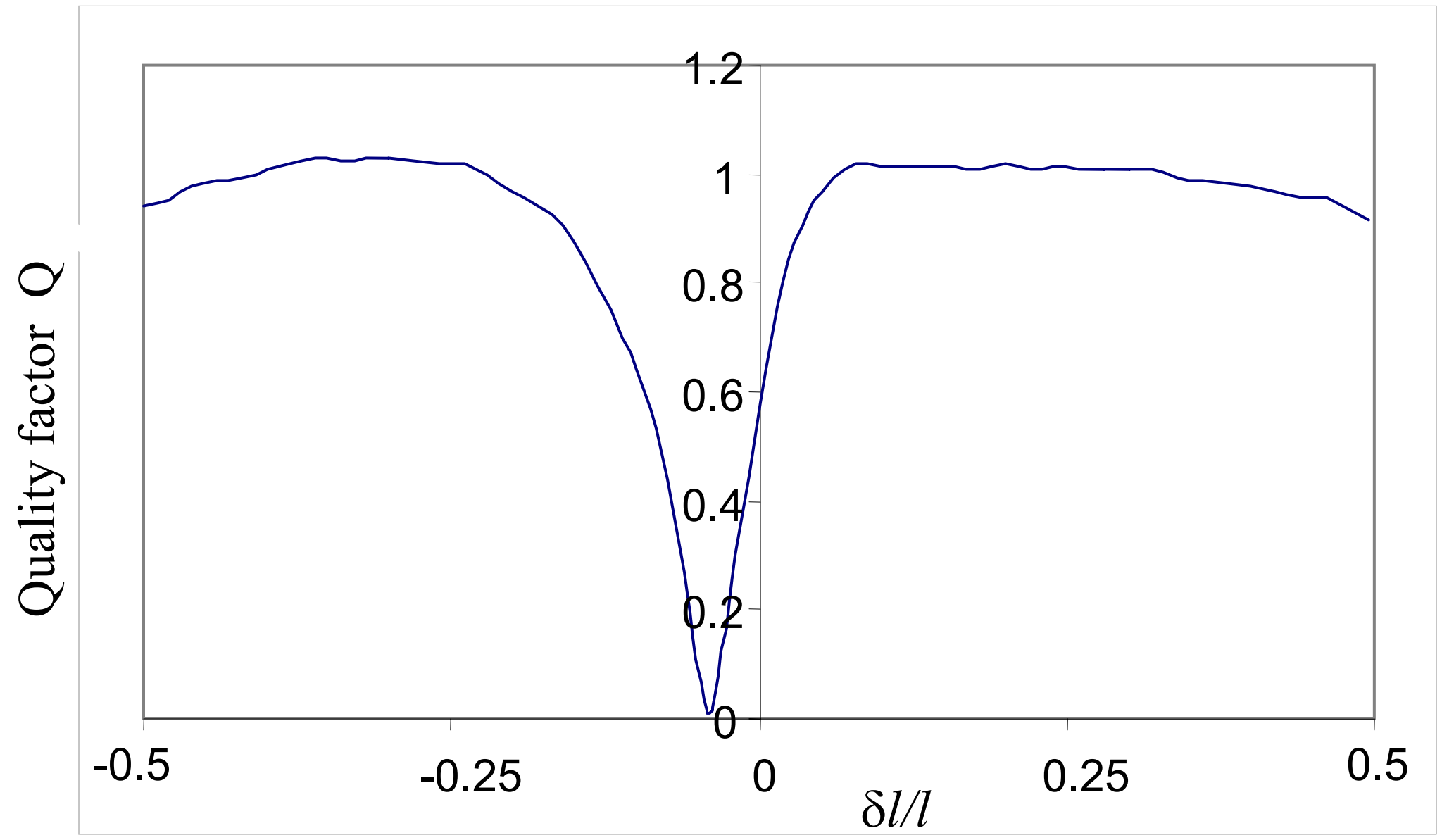

FIG. 14. Quality Factor Q vs. $\delta l/l$ for the Continuous Interferometer, simulated with parameters $\Omega_0 = 2\pi(7 \times 10^4)$, $L = 3 \times 10^{-3}$ m., $\Omega_0 T = 3.3$.

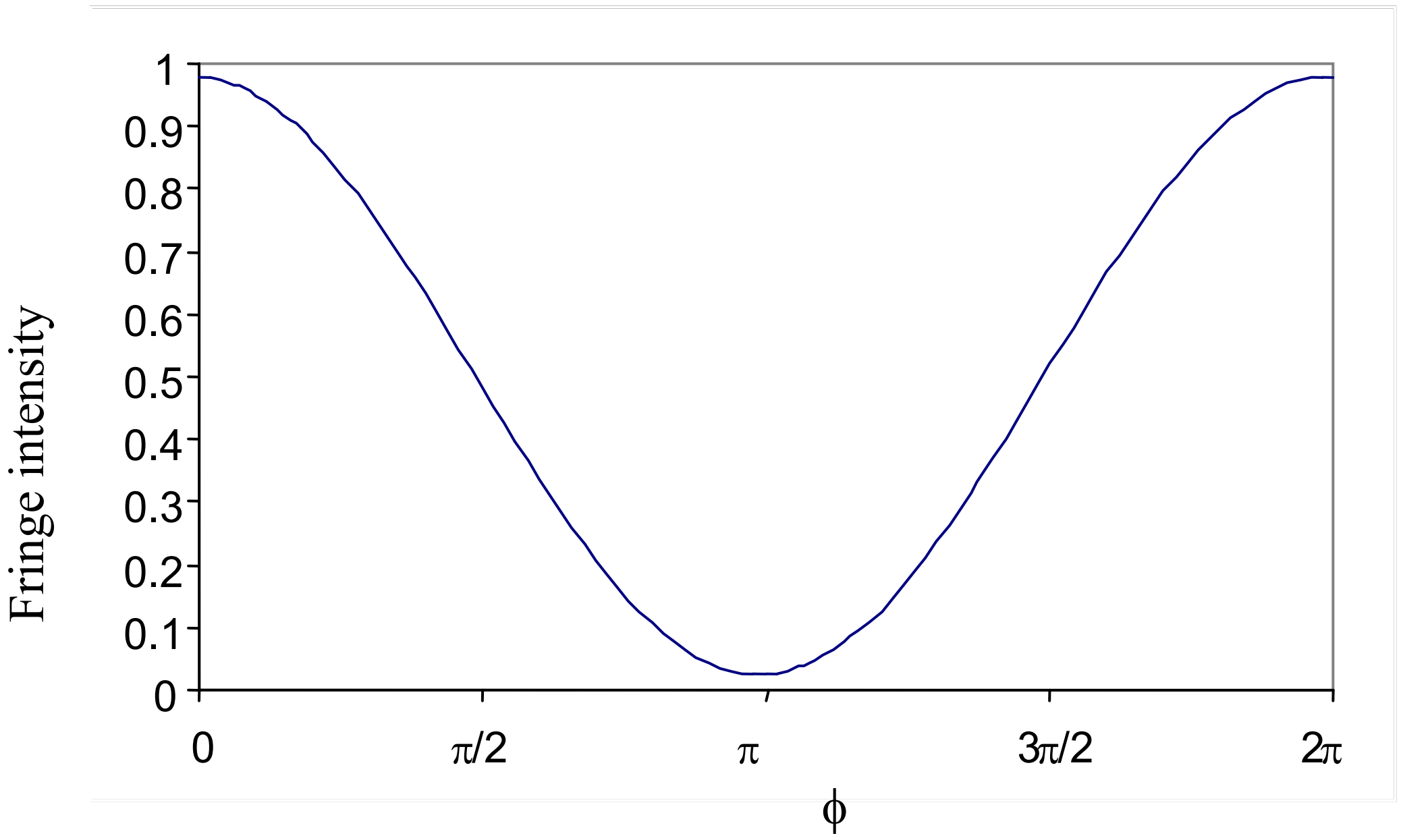

FIG. 15. Phase scan for the Continuous Interferometer when the phase is applied at $\delta l/l = \pm 12/25$, giving the maximum fringe contrast most comparable to the BCI. Simulated with parameters $\Omega_0 = 2\pi(7 \times 10^4)$, $L = 3 \times 10^{-3}$ m., v=300m/s, $\Omega_0 T = 3.3$.

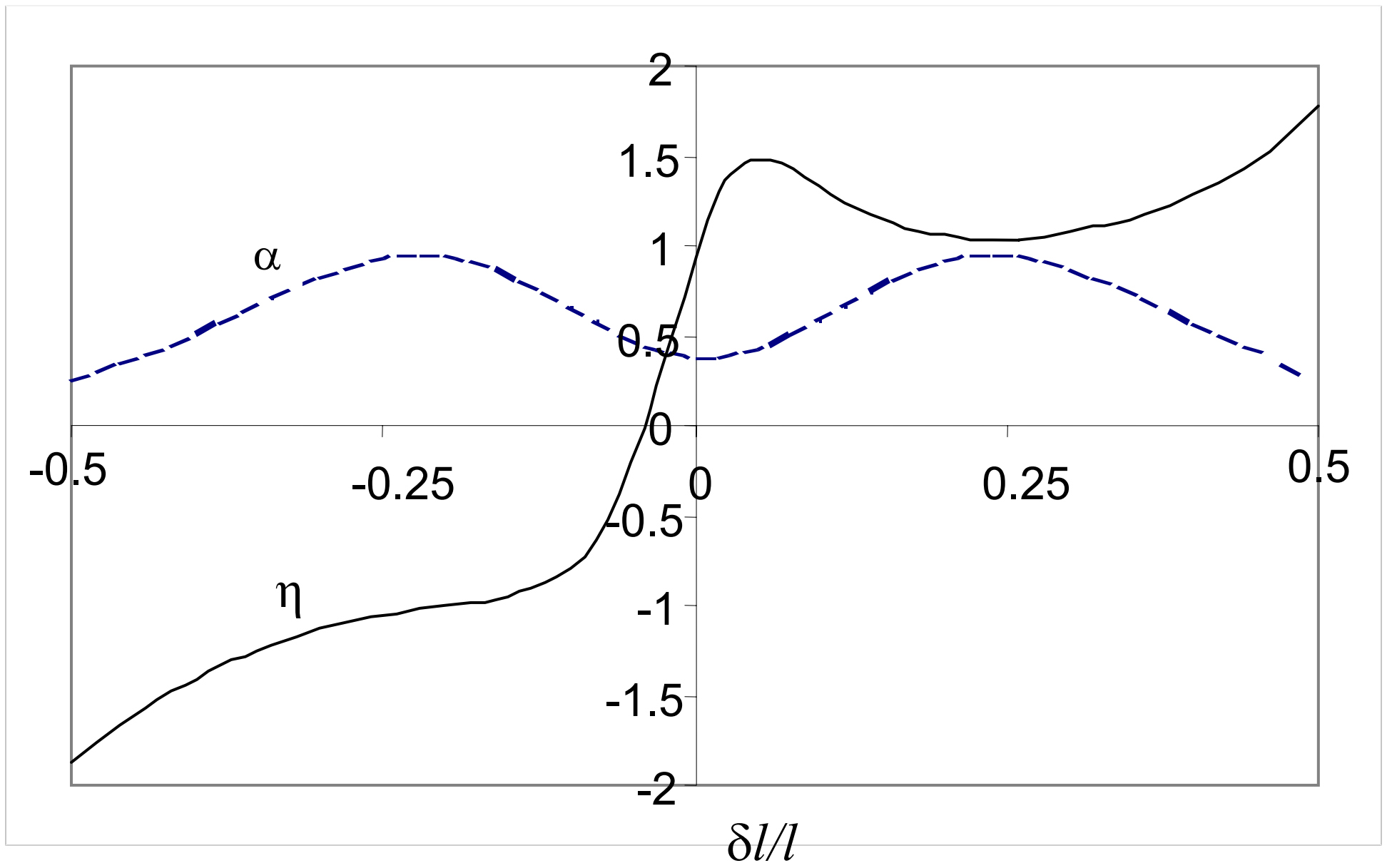

FIG. 16. $\eta$ ratio (solid line) and signal amplitude $\alpha$ (dashed line) vs. $\delta l/l$ for the Continuous Interferometer, simulated with parameters $\Omega_0 = 2\pi(7 \times 10^4)$, $L = 3 \times 10^{-3}$ m., $\Omega_0 T = 3.3$. The signal amplitude is symmetric around $\delta l = 0$, and reaches a maximum at $\delta l = \pm\ 12/25$. At these values the interferometer behaves most like a BCI, with the effective area also leveling off temporarily before increasing.

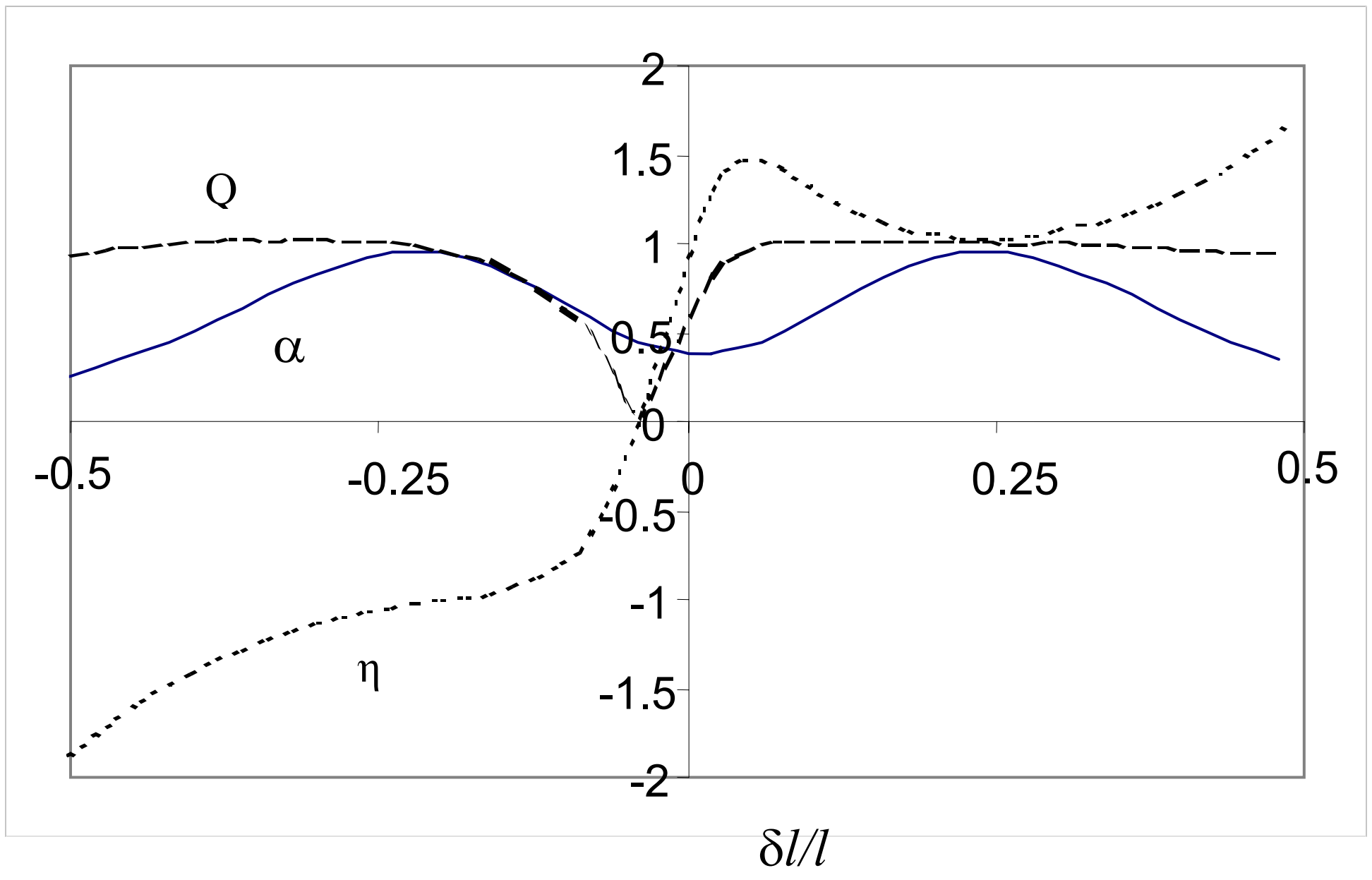

FIG. 17. Normalized values of the effective area η (short dashed line), the quality factor Q (long dashed line), and the signal amplitude α (solid line) vs. $\delta l/l$ for the Continuous Interferometer, simulated with parameters $\Omega_0 = 2\pi(7 \times 10^4)$, $L = 3 \times 10^{-3}$ m., $\Omega_0 T = 3.3$. This shows that the quality factor is very similar to that of the modified BCI. If the phase is applied starting away from $\delta l/l = -2/25$, Q is approximately 1, which means that the performance of this interferometer is comparable to that of a BCI.